\documentclass[a4paper,fleqn,usenatbib]{mnras}

\usepackage{mathptmx}

\usepackage[T1]{fontenc}
\usepackage{ae,aecompl}

\usepackage{graphicx}	
\usepackage{amsmath}	
\usepackage{amssymb}	

\title[Dynamics of multiple populations]{Dynamical Evolution of Multiple-Population Globular Clusters}

\author[E. Vesperini et al.]{
Enrico Vesperini,$^{1}$\thanks{E-mail: evesperi@indiana.edu}
Jongsuk Hong,$^{2,3}$
Mirek Giersz,$^{4}$
Arkadiusz Hypki$^{4}$
\\
$^{1}$Department of Astronomy, Indiana University, Bloomington, Swain West, 727 E. 3rd Street, IN, 47405, USA\\
$^{2}$Department of Astronomy, Yonsei University 50 Yonsei-Ro, Seodaemun-Gu, Seoul 03722, Republic of Korea\\
$^{3}$Korea Astronomy and Space Science Institute, Daejeon 34055, Republic of Korea\\
$^{4}$Nicolaus Copernicus Astronomical Centre, Polish Academy of Sciences, ul. Bartycka 18, 00-716 Warsaw, Poland\\
}

\pubyear{2020}

\begin{document}
\newcommand{\msg}{$M_{\rm 2G}/M_{\rm tot}$}
\newcommand{\msgsp}{$M_{\rm 2G}/M_{\rm tot}~$ }
\newcommand{\msun}{$m_{\odot}$}
\newcommand{\rhratio}{$r_{\rm h,1G}/r_{\rm h,2G}$}
\newcommand{\stsr}{$\sigma_T/\sigma_R-1$}
\newcommand{\escnsg}{$(N_{\rm 2G}/N_{\rm tot})_{\rm esc}$}
\label{firstpage}
\pagerange{\pageref{firstpage}--\pageref{lastpage}}
\maketitle

\begin{abstract}
We have carried out a set of Monte Carlo simulations to study a number of fundamental aspects of the dynamical evolution of multiple stellar populations in globular clusters with different initial masses, fractions of second generation (2G) stars, and structural properties. Our simulations explore and elucidate: 1) the role of early and long-term dynamical processes and stellar escape in the evolution of the fraction of 2G stars and the link between the evolution of the fraction of 2G stars and various dynamical parameters; 2) the link between the fraction of 2G stars inside the cluster and in the population of escaping stars during a cluster's dynamical evolution; 3) the dynamics of the spatial mixing of the first-generation (1G) and 2G stars and the details of the structural properties of the two populations as they evolve toward mixing; 4) the implications of the initial differences between the spatial distribution of 1G and 2G stars for the evolution of the anisotropy in the velocity distribution and the expected radial profile of the 1G and 2G anisotropy for clusters at different stages of their dynamical history; 5) the variation of the degree of energy equipartition of the 1G and the 2G populations as a function of the distance from the cluster's centre and the cluster's evolutionary phase.
\end{abstract}




\section{Introduction}
Globular clusters host multiple stellar populations characterized by a variety of patterns in the abundances of a number of light elements (e.g. C, N, Na, O, Mg, Al); a few clusters also show further complexities in the chemical properties of their populations including variations in the abundances of Fe and s-process elements (see e.g. Gratton et al. 2019 for a recent review and references therein).
The multiple stellar populations found in many clusters differ not only in their chemical abundances but also in their structural and kinematic properties (see e.g. Sollima et al. 2007, Bellini et al. 2009, Lardo et al. 2011, Simioni et al. 2016, Dalessandro et al. 2019, Richer et al. 2013, Bellini et al. 2015, 2018 Cordero et al. 2017, Lee 2017, Cordoni et al. 2020a, 2020b); these differences  clearly reveal the existence of a tight link between the origin of the populations' chemical properties and the cluster's formation and dynamical history and add another fundamental node in the complex network of relationships between dynamics and the properties of the stellar content of globular clusters.

The numerous observational and theoretical studies of multiple stellar populations are revealing  an increasingly complex picture of the chemical and dynamical properties of globular clusters and raising many new questions.
The main issues in the study of multiple populations in globular clusters concern the origin of the processed gas from which stars with anomalous chemical properties form, the formation history of multiple populations, the differences in their initial structural and kinematic properties, how these differences affected the cluster dynamical evolution, and the extent to which the current properties retain memory of these initial differences.

As for the gas out of which second generation stars formed  (hereafter 2G; notice that according to some of the models listed below different populations would form at the same time and in that case it is more appropriate to use the terms first- and second-population to denote stars with different chemical properties)  a variety of different stellar sources have been proposed including AGB stars, rapidly-rotating massive stars and massive binaries, supermassive stars, black hole accretion disks, stellar mergers (see e.g. Ventura et al. 2001, Decressin et al. 2007, de Mink et al. 2009, Bastian et al. 2013, Krause et al. 2013, D'Ercole et al. 2008, 2010, 2012, Bekki 2010, Bekki et al. 2017, Denissenkov \& Hartwick 2014, D'Antona et al. 2016, Elmegreen 2017, Kim \& Lee 2018, Breen 2018, Gieles et al. 2018,  Calura et al. 2019, Howard et al. 2019, Wang et al. 2020, McKenzie \& Bekki 2021). No consensus has been reached on this fundamental aspect of the study of multiple populations and many of these models still require significant further development  concerning the nucleosynthesis of the proposed polluters and the comparison with all the observational constraints (see Gratton et al. 2019).

As pointed out above, different stellar populations have been found in many cases to be characterized by different dynamical properties and the study of the dynamical implications of these differences is essential to build a complete picture of globular clusters' formation and evolution. 

In all models dynamics plays a key role in determining the initial differences in the structural and kinematic properties of multiple populations and their subsequent evolution (see e.g. D'Ercole et al. 2008, Bekki 2010, 2011, Vesperini et al. 2013, Calura et al. 2019), and, in some of the models presented in the literature, the cluster's dynamics plays a central role also in the formation of the stellar polluters providing gas for the formation of 2G stars (see e.g. Gieles et al. 2018, Wang et al. 2020). 

The first studies of multiple-population cluster dynamics have addressed a number of aspects related to the hydrodynamics of the  gas out of which 2G stars formed (see e.g. D'Ercole et al. 2008, 2016, Bekki 2010, 2011, Calura et al. 2019). These studies have shown that 2G stars tend to form spatially concentrated in the cluster's inner regions; other models, although not based on detailed hydrodynamical simulations, also agree and provide qualitative considerations supporting this expected difference between the initial spatial distributions of 2G and 1G stars.
A few previous studies (see e.g. Vesperini et al. 2013, 2018, Miholics et al. 2015, Henault-Brunet et al. 2015, Fare et al. 2018, Tiongco et al. 2019) have shown that initial differences in the spatial distribution of 1G and 2G stars are gradually erased during a cluster's evolution. Dynamically younger clusters may retain some memory of the initial differences until the present day while in clusters that have reached the more advanced stages of their evolution the two populations are expected to be completely mixed.
Examples of both of these cases have been found in observational studies (see e.g. Sollima et al. 2007, Bellini et al. 2009, Lardo et al. 2011, Beccari et al. 2013, Cordero et al. 2014, Simioni et al. 2016, Boberg et al. 2016, Lee 2017, Gerber et al. 2020b for studies showing clusters in which the 2G is more concentrated than the 1G and Dalessandro et al. 2014, Nardiello et al. 2015, Cordero et al. 2015, Gerber et al. 2018, 2020a for clusters in which the two populations are instead completely mixed) and a recent analysis by Dalessandro et al. (2019) has provided an observational picture of the evolutionary path toward spatial mixing.

Dynamical differences between the 2G and the 1G populations are not limited to their structural properties: a number of studies have shown that, either as a result of the formation process or  the subsequent dynamical evolution, the kinematic properties of the 2G and 1G stars may differ: the 2G population may be characterized by a more rapid rotation and a more radially anisotropic velocity distribution than 1G stars (see e.g. Bekki 2010, Mastrobuono-Battisti \& Perets 2013, 2016, Bellini et al. 2015, Henault-Brunet et al. 2015, Tiongco et al. 2019). The first observational studies of the kinematics of multiple populations have revealed such differences in a few clusters (see e.g. Richer et al. 2013, Bellini et al. 2015, 2018, Cordero et al. 2017, Milone et al. 2018, Libralato et al. 2019, Cordoni et al. 2020a, 2020b; see also e.g. Pancino et al. 2007, Cordoni et al. 2020b, for clusters in which the multiple populations share similar kinematic properties).
Significant additional efforts beyond these initial pioneering studies will be necessary to draw a more complete observational picture of the dynamical properties of multiple-population clusters. In particular, it will be necessary to extend the observational studies to cover a wider radial range of distances from clusters' centres to include the outer regions where structural and kinematic differences, if not completely erased during a cluster's long-term dynamical  evolution, should be stronger.

The differences between the 1G and the 2G structural properties have also been shown to have important implications for the evolution and survival of binary stars. For the more centrally concentrated 2G population, binaries may be disrupted and evolve more rapidly than 1G binaries leading to different 1G and 2G binary fractions and differences between the orbital properties of the surviving binaries (see e.g. Vesperini et al. 2011, Hong et al. 2015, 2016, 2019, Lucatello et al. 2015, Dalessandro et al. 2018, Gratton et al. 2019, Milone et al. 2020, Kamann et al. 2020 for some theoretical and observational studies on the properties of binaries in multiple populations). The possible role of the differences between the formation history  of 1G and 2G populations on the formation of binaries and their initial properties is still unexplored and is certainly another important aspect in the study of binaries in multiple-population clusters.

In this paper, we present the results of a set   Monte Carlo simulations following the evolution of multiple-population globular clusters. We explore the role played by early and long-term evolutionary processes in the dynamical evolution of multiple populations, in driving the escape from the cluster of 1G and 2G stars, and  determining the evolution of the fraction of 2G stars in the cluster. We follow the evolution of the main structural and kinematic properties during the various stages of a cluster's evolution and discuss  what fingerprints of the formation process may still be observable today along with the dynamical signatures and implications of the initial differences between the structural properties of the 1G and the 2G populations.

This is the structure of the paper: in Section 2 we describe the method and the initial conditions used for our simulations. In section 3 we present our results, and we summarize our conclusions in section 4.

\section{Methods and Initial Conditions}
\label{sec:methods}
This study is based on a set of Monte Carlo simulations run with the MOCCA code (Hypki \& Giersz 2013, Giersz et al. 2013). The code includes the effects stellar and binary evolution modeled using the SSE and BSE codes (Hurley et al. 2000,2002) and assuming supernovae kick velocities folllowing a Mawellian distribution with a dispersion equal to 265 km/s (Hobbs et al. 2005), two-body relaxation and binary star interactions, and a truncation radius mimicking the effect of the tidal truncation due to the external tidal field of the host galaxy for a cluster on a circular orbit (see Hypki \& Giersz 2013, Giersz et al. 2013 for further details on the MOCCA code).
\begin{table*}
	\caption{Summary of initial conditions}
	\label{tab:tab1}
	\begin{tabular}{lllllll} 
		\hline
		id. & N & $M_{\rm 2G}/M_{\rm tot}$ & $r_{\rm h,1G}/r_{\rm h,2G}$& $(W_{0,1G}, W_{0,2G})$ & $r_{\rm h}/r_{\rm tidal}$& $r_{\rm tidal}$ \\
		\hline
		sg01c20sf & $10^6$ & 0.1 & 20 & (5,7) & 0.17& 77\\
		sg01c20wf & $10^6$ & 0.1 & 20 & (5,7) & 0.17& 122.3\\
		sg025c20sf & $10^6$ & 0.25 & 20 & (5,7) & 0.14& 77\\
		sg025c20wf & $10^6$ & 0.25 & 20 & (5,7) & 0.14& 122.3\\
		sg025c20w04sf & $10^6$ & 0.25 & 20 & (4,7) & 0.17& 77\\
		sg025c20w04wf & $10^6$ & 0.25 & 20 & (4,7) & 0.17& 122.3\\
		sg025c10sf & $10^6$ & 0.25 & 10 & (5,7) & 0.14& 77\\
		sg025c10wf & $10^6$ & 0.25 & 10 & (5,7) & 0.14& 122.3\\
		sg04c20sf & $10^6$ & 0.4 & 20 & (5,7) & 0.09&77\\
		sg04c20wf & $10^6$ & 0.4 & 20 & (5,7) & 0.09&122.3\\
		sg04c10sf & $10^6$ & 0.4 & 10 & (5,7) & 0.095&77\\
		sg04c10wf & $10^6$ & 0.4 & 10 & (5,7) & 0.095&122.3\\
		4m-sg01c20sf & $4 \times 10^6$ & 0.1 & 20 & (5,7) & 0.17&122.2\\
		4m-sg01c20wf & $4 \times 10^6$ & 0.1 & 20 & (5,7) & 0.17&193.9\\
		4m-sg025c20sf & $4 \times 10^6$ & 0.25 & 20 & (5,7) & 0.14&122.2\\
		4m-sg025c20wf & $4 \times 10^6$ & 0.25 & 20 & (5,7) & 0.14&193.9\\
		4m-sg025c10sf & $4 \times 10^6$ & 0.25 & 10 & (5,7) & 0.14&122.2\\
		4m-sg025c10wf & $4 \times 10^6$ & 0.25 & 10 & (5,7) & 0.14&193.9\\
		\hline
	\end{tabular}
\end{table*}

Table 1 summarizes the initial conditions of all our simulations along with the id used throughout the paper to refer to each simulation.

The systems we have considered for our study start with a total number of stars equal to $N=10^6$ or $N=4 \times 10^6$ with no primordial binaries and with masses distributed according to a Kroupa (2001) initial mass function (IMF) between 0.1 $m_{\odot}$ and 100 $m_{\odot}$. In all the systems, the density profile of 2G stars is modeled as a King model with $W_0=7$ (where $W_0$ is the central dimensionless potential; King 1966) while the 1G population is less concentrated and distributed according to a tidally limited King model with  $W_0=4$ or $W_0=5$. In all cases the 2G population is confined in the inner regions of the cluster; we have explored models with values of the initial ratio of the 1G half-mass radius to the  2G half-mass radius, $r_{\rm h,1G}/r_{\rm h,2G}$, equal to 20 and 10.
As for the initial ratio of the total 2G mass to the total cluster mass, \msg, we have studied systems with \msg=0.1, 0.25 and, for a few models, 0.4. The initial conditions adopted follow the general trend emerging from the results of a number of hydro simulations of multiple population formation (see e.g. D'Ercole et al. 2008, Bekki 2010, 2011, Calura et al. 2019) showing that 2G stars form centrally concentrated in a more extended 1G system. As for the initial 2G mass fraction, we have explored a few different values since this quantity is still poorly constrained and may depend on a number of factors including the amount of gas supplied by the 1G stellar polluters and the amount of pristine gas involved in the 2G star formation (see e.g. Calura et al. 2019).

All the systems start with an isotropic velocity distribution but for the model sg025c20wf we have also explored two initial conditions with an initially radially anisotropic velocity distribution with anisotropy following an Osipkov-Merritt profile (see e.g. Binney \& Tremaine 2008), $\beta=1-(\sigma_{\theta}^{2}+\sigma_{\phi}^{2})/(2\sigma_{r}^2)=1/(1+r_a^2/r^2)$ where $\sigma_{\theta}$, $\sigma_{\phi}$, and $\sigma_r$ are the three components of the velocity dispersion in spherical coordinates, and $r_a$ is an anisotropy scale radius beyond which the velocity distribution becomes increasingly radaially anisotropic. For the two anisotropic models we have studied, the anisotropy radius, $r_a$ was set equal to  $r_{\rm h}/2$ and $r_{\rm h}/3$ where $r_{\rm h}$ is the cluster's initial half-mass radius.

Finally we have considered two different values for the initial truncation radius corresponding to tidal fields with different strength which we will refer to as strong field (sf) and weak field (wf). The two values of the truncation radii adopted would correspond to the tidal radii of clusters with the masses of our models  and moving on  circular orbits in a logarithmic potential at galactocentric distances of 4 kpc and 8 kpc for the strong and the weak field case respectively.

We emphasize that the set of initial conditions explored here are not meant to provide a complete coverage of the possible range of the various structural and kinematic properties, and other parameters characterizing multiple-population clusters. The goal of this paper is to study the fundamental aspects of the early and long-term evolution of multiple-population clusters, and illustrate the role of various processes in the dynamics of multiple populations.  An extension of the set of simulations presented here is currently in progress. In the extended suite of simulations, besides considering  a broader range of initial structural properties, we will include a population of primordial binaries. Although primordial binaries are not expected to significantly affect the general dynamical aspects studied in this paper, as discussed in the Introduction the study of the differences in the evolution and survival of 1G and 2G binaries can provide a key additional dynamical insight into the formation and dynamical of multiple-population clusters. Another important aspect that will be further explored in a future study concerns the evolution of rotating multiple-population clusters. The Monte Carlo code used in this study can not follow the evolution of rotating stellar systems and the study of rotating systems will be pursued by N-body simulations (see e.g. Mastrobuono-Battisti \& Perets 2013, 2016, Henault-Brunet et al. 2015, Tiongco et al. 2019 for some early studies of rotating multiple-population clusters). The planned extensions of the study presented here will allow us to further explore the origin of the trends observed in the properties of multiple populations in the Galactic globular cluster system and their possible connection with specific range of initial properties set by the formation processes.

\section{Results}

\subsection{Evolution of the fraction of second-generation stars}
\label{sec:sgfraction}
Many   formation models predict that the fraction of the total cluster's mass in 2G stars, \msg, was initially smaller than its current value and increased to reach its present value as a result of the preferential loss of 1G stars during the cluster's dynamical evolution.
Here we explore the possible dynamical path and processes behind the evolution  of \msg.

\begin{figure}
	\includegraphics[width=\columnwidth]{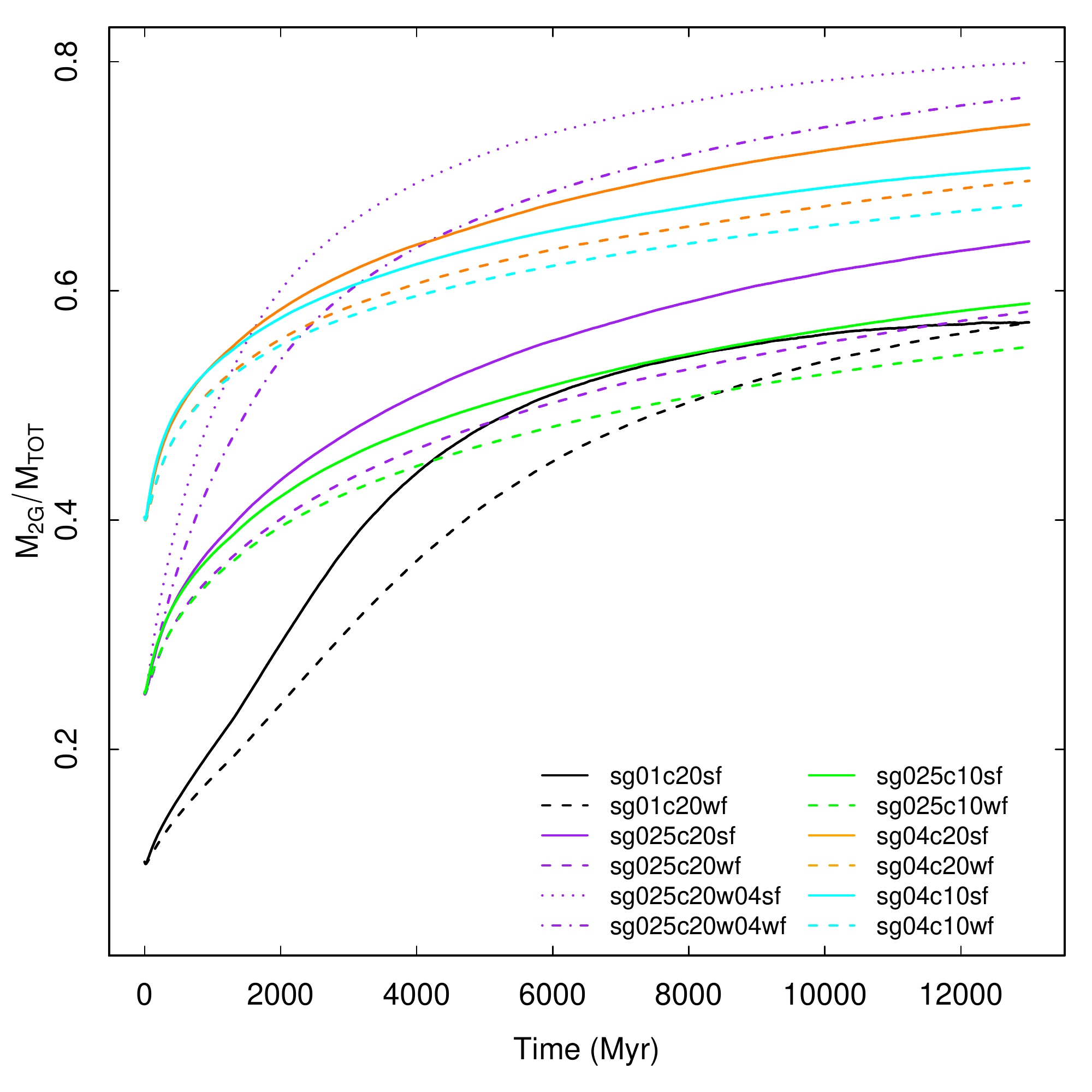}
    \caption{Time evolution of the ratio of the total mass in 2G stars to the total mass of the cluster for models starting with $10^6$ stars: sg01c20sf (solid black line), sg01c20wf (dashed black line), sg025c20sf (solid purple line), sg025c20wf (dashed purple line), sg025c20w04sf (dotted purple line), sg025c20w04wf (dot-dashed purple line), sg025c10sf (solid green line), sg025c10wf (dashed green line), sg04c20sf (solid orange line), sg04c20wf (dashed orange line), sg04c10sf (solid cyan line), sg04c10wf (dashed cyan line).}
    \label{fig:msgmtot1m}
\end{figure}

\begin{figure}
	\includegraphics[width=\columnwidth]{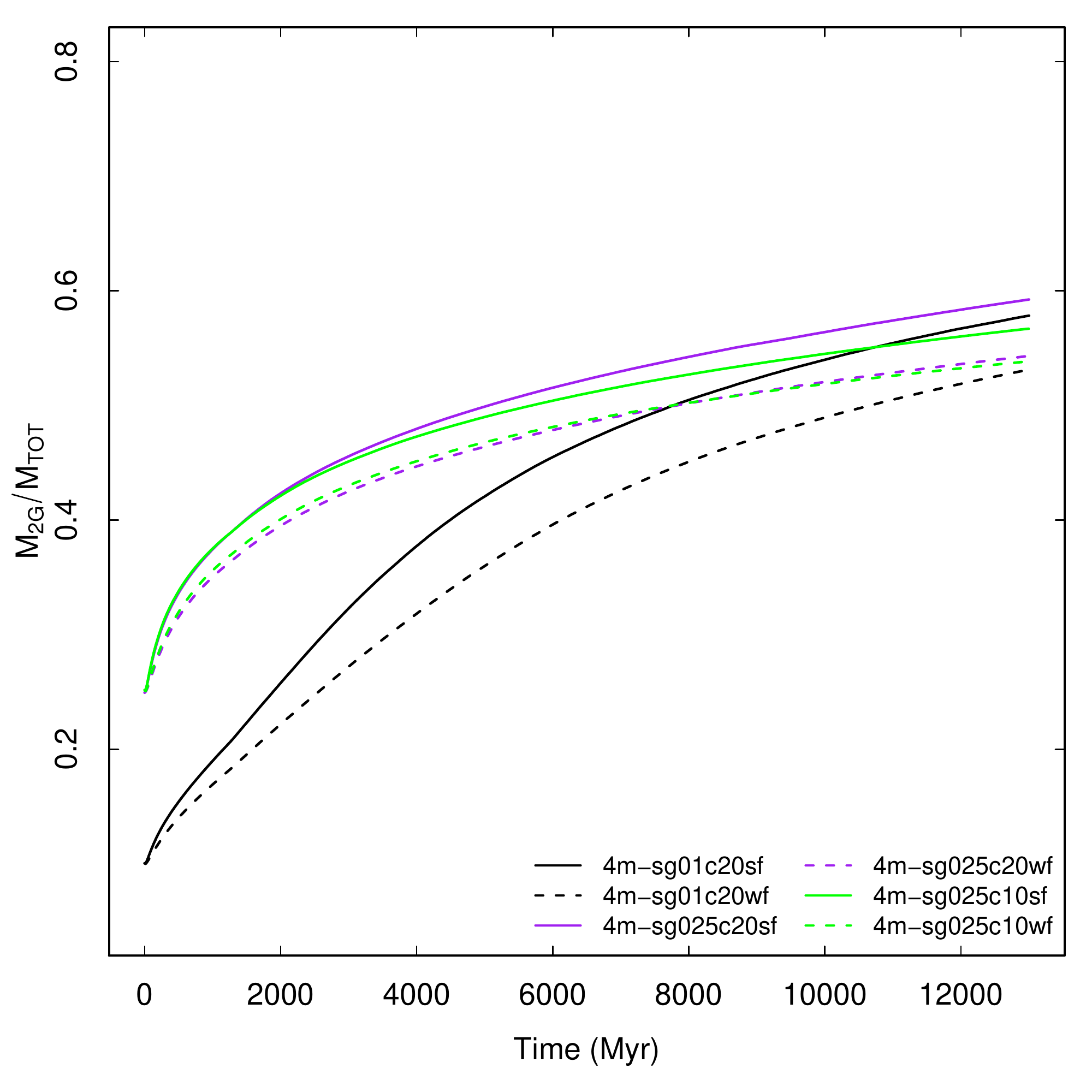}
    \caption{Time evolution of the ratio of the total mass in 2G stars to the total mass of the cluster for models starting with $4\times 10^6$ stars: 4m-sg01c20sf (solid black line), 4m-sg01c20wf (dashed black line), 4m-sg025c20sf (solid purple line), 4m-sg025c20wf (dashed purple line), 4m-sg025c10sf (solid green line), 4m-sg025c10wf (dashed green line).}
    \label{fig:msgmtot4m}
\end{figure}

In Figs. \ref{fig:msgmtot1m} and \ref{fig:msgmtot4m} we show the time evolution of \msgsp for a few representative models starting with a range of different initial values of \msg, different tidal radii, and different initial cluster's masses and structural parameters.

\begin{figure}
	\includegraphics[width=\columnwidth]{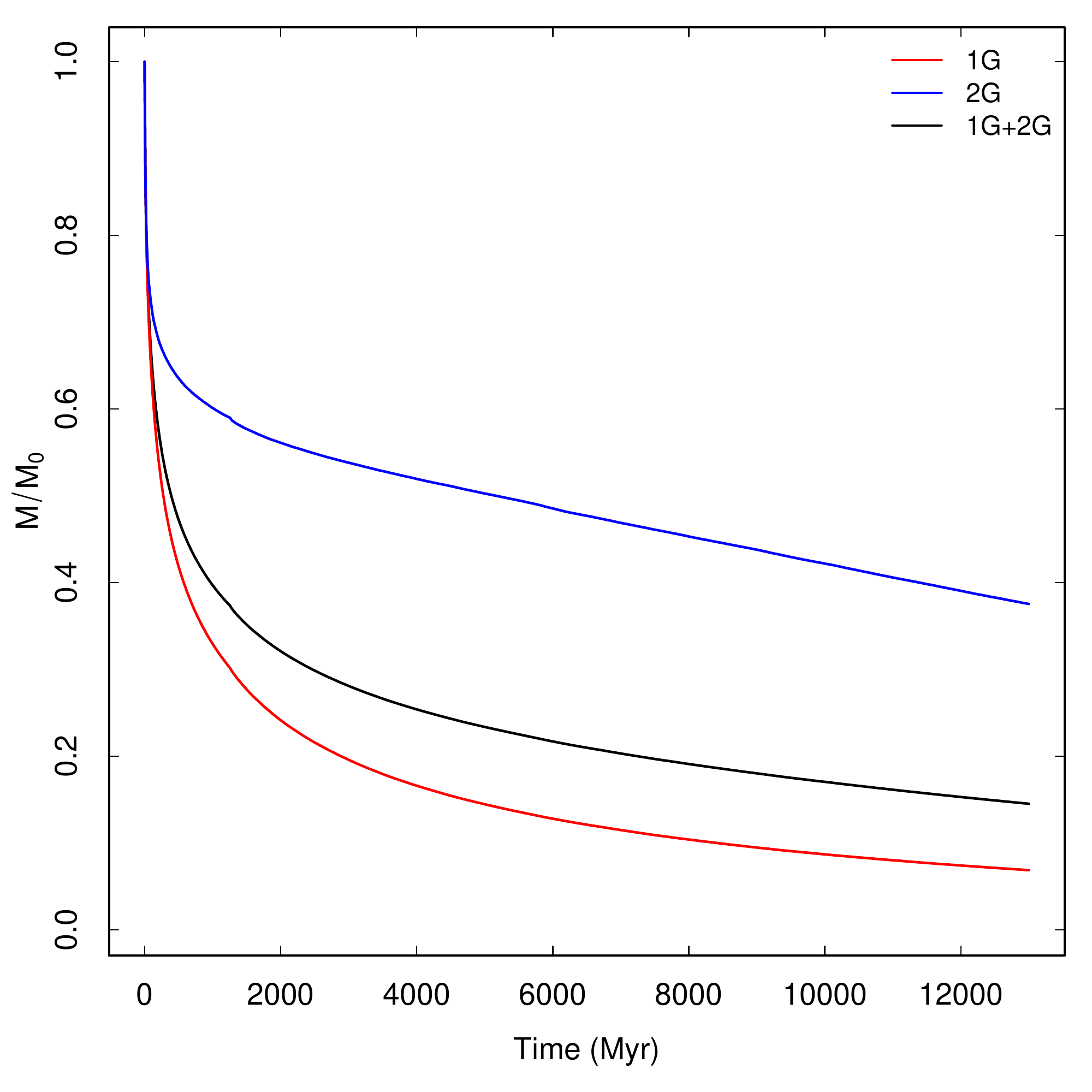}
    \caption{Time evolution of the total mass of 1G stars (red line),  2G stars (blue line), and of the total cluster's mass (black line) each normalized to their respective initial values for the sg025c20sf model. }
    \label{fig:mvstime}
\end{figure}

\begin{figure}
	\includegraphics[width=\columnwidth]{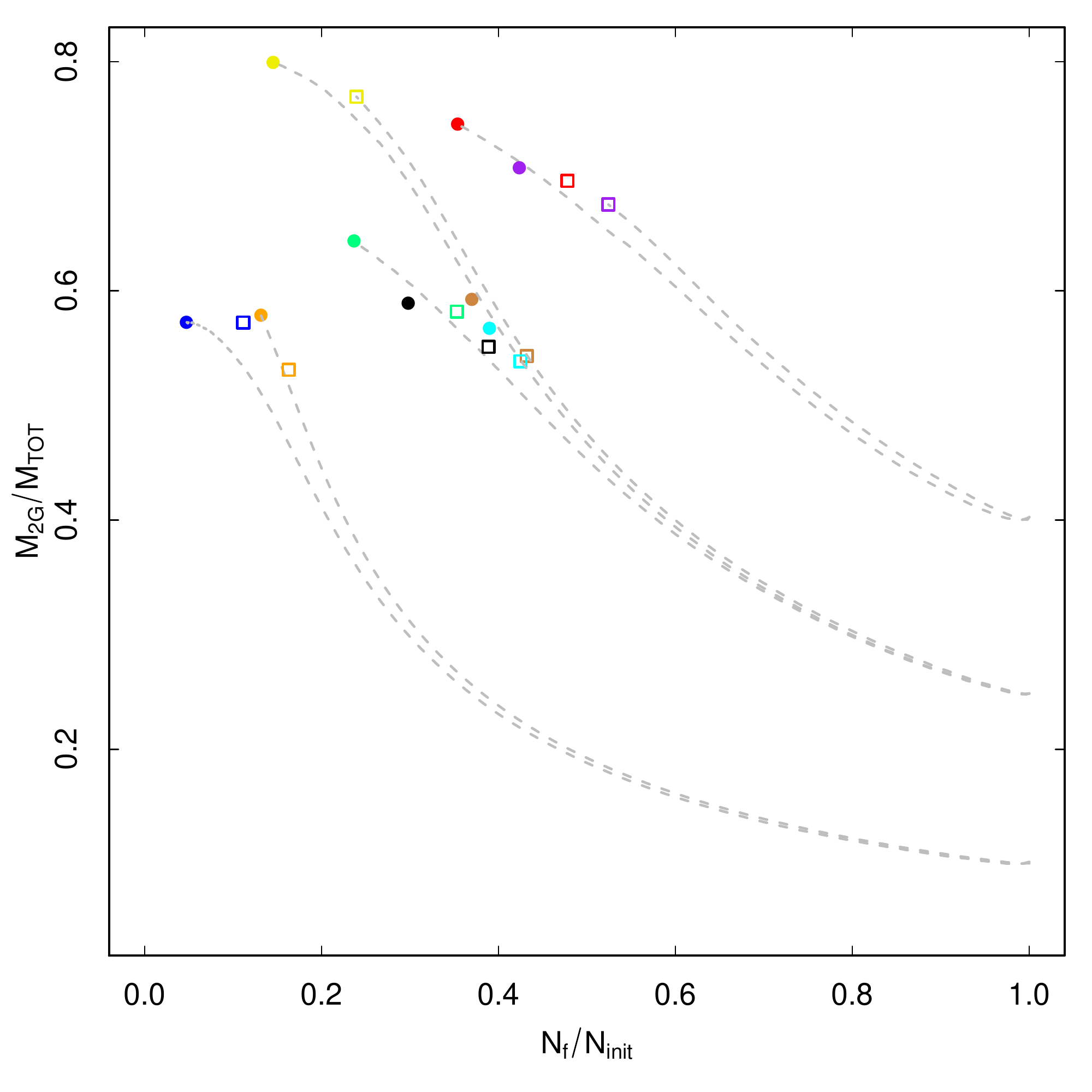}
        \caption{Final values of \msgsp versus the ratio of the final to the initial number of stars: 4m-sg01c20sf (orange filled dot), 4m-sg01c20wf (orange square), 4m-sg025c20sf (brown filled dot), 4m-sg025c20wf (brown square), 4m-sg025c10sf (cyan filled dot), 4m-sg025c10sf (cyan square), sg01c20sf (blue filled dot), sg01c20wf (blue square), sg025c20sf (green filled dot), sg025c20wf (green square), sg025c20w04sf (filled yellow dot), sg025c20w04wf (yellow square), sg025c10sf (black filled dot), sg025c10wf (black square), sg04c20sf (red filled dot), sg04c20wf (red square), sg04c10sf (purple filled dot), sg04c10wf (purple square). Dashed lines show the full evolutionary tracks for a few representative models.}
    \label{fig:msgvsnfni}
\end{figure}

In all of our models most of the increase in \msgsp occurs during the first few Gyrs of a cluster's evolution. During this early evolutionary phase the effects of two-body relaxation are negligible and the preferential loss of 1G stars leading to the early increase of \msgsp is driven by the cluster's expansion triggered by mass loss due to stellar evolution. This early expansion and the dynamical effects induced by this mass loss have been thouroghly investigated in the context of single-population clusters in several early pioneering studies (see e.g. Chernoff \& Shapiro 1987, Chernoff \& Weinberg 1990, Fukushige \& Heggie 1995) which have shown that this process can cause a cluster to lose a significant number of stars and even undergo an early complete dissolution (see also Vesperini et al. 2009, Whitehead et al. 2013, Contenta et al. 2015, Giersz et al. 2019 for more recent studies further exploring different aspects of this path to cluster dissolution). In the context of the evolution of multiple-population clusters we have discussed the possible role of this dynamical process for the evolution of the fraction of 2G stars in our early simulations presented in D'Ercole et al. (2008). The Monte Carlo simulations presented in this paper allow us to carry out a significantly more detailed characterization of this early dynamical phase and a more extensive study of its dependence on a number of initial parameters.

As shown in Figs. \ref{fig:msgmtot1m} and \ref{fig:msgmtot4m}, the early expansion triggered by stellar evolution mass loss affects primarily the more extended 1G population while the centrally concentrated spatial distribution of the 2G subsystem prevents any significant loss of 2G stars.  After the first few Gyrs of evolution, clusters enter a dynamical phase driven by the effects of two-body relaxation; during this phase the loss of stars due to two-body relaxation affects both 1G and 2G stars with a slight preferential loss of 1G stars causing \msgsp to continue to slightly increase. In all the cases investigated the final values of \msgsp fall within the range of  those found in observational studies showing that for Galactic clusters \msg $\sim 0.35-0.9$ (see Milone et al. 2017). 

The results of our simulations also clearly illustrate the dependence of the evolution of \msgsp on the structural properties of the 1G system. For models in which the 1G starts with a lower concentration density profile (models sg025c20w04sf and sg025c20w04wf), the initial increase in \msgsp is more rapid and substantial as the early expansion of less concentrated 1G systems leads to a larger loss of 1G stars (see the purple dotted and dot-dashed lines in Fig. \ref{fig:msgmtot1m} for the  evolution of the sg025c20w04sf and sg025c20w04wf models).

In Fig. \ref{fig:mvstime},  we show the time evolution of the total masses of the 1G and 2G systems along with the total cluster's mass (each normalized to their respective initial values) for one of our models. The model shown in this figure is one of those significantly affected by both early and long-term mass loss and serves to illustrate the early loss of stars suffered by the 1G system and the transition from the early evolutionary stages dominated by the expansion triggered by mass loss due to stellar evolution (and the loss of stars that ensues from this expansion) to the long-term evolution driven by two-body relaxation.

\begin{table}
	\caption{Final number of stars and 2G mass fraction}
	\label{tab:tab2}
	\begin{tabular}{lcc} 
		\hline
		id. & $N_f$ & $M_{\rm 2G}/M_{\rm tot}$ \\
		\hline
		sg01c20sf & 47,381  & 0.57 \\
		sg01c20wf & 111,517 & 0.57 \\
		sg025c20sf & 236,731 & 0.64 \\
		sg025c20wf & 352,947 & 0.58 \\ 
		sg025c20w04sf & 145,363 & 0.80 \\ 
		sg025c20w04wf & 239,498  & 0.77 \\
		sg025c10sf & 297,975 &  0.59 \\
		sg025c10wf & 388,698 &  0.55 \\
		sg04c20sf & 353,817 &   0.75 \\
		sg04c20wf & 477,971 &  0.70 \\
		sg04c10sf & 423,635 &  0.71\\
		sg04c10wf & 524,268 &  0.68\\
		4m-sg01c20sf & 526,262 & 0.58\\
		4m-sg01c20wf & 651,973 & 0.53\\
		4m-sg025c20sf & 1,479,567  & 0.59\\
		4m-sg025c20wf & 1,727,911  & 0.54\\
		4m-sg025c10sf & 1,559,549  & 0.57\\
		4m-sg025c10wf & 1,699,293  & 0.54\\
		\hline
	\end{tabular}
\end{table}

The link between the evolution of \msgsp and stellar escape is further illustrated  in Fig. \ref{fig:msgvsnfni}: this figure shows the final values of \msgsp versus the ratio of the final to the initial number of stars.  Notice that, as expected, models evolving in a strong field are characterized by smaller values of $N_f/N_i$; this is a consequence of the stronger loss of stars due to the effects of two-body relaxation during the clusters' long-term evolution. We report in Table  \ref{tab:tab2} the final values (at $t=13$ Gyr) of the total number of stars, $N_f$, and \msg.

As pointed out above, in our models, the loss of stars due to the effects of two-body relaxation during the cluster's long-term evolution causes only a slight additional increase in \msgsp compared to the more significant increase occurring during the cluster's early evolution. The dominant role played by early evolutionary processes in determining the current values of \msgsp implies that no significant dependence of \msgsp on the galactocentric distance is to be expected since, for the clusters with the tidally truncated structural properties adopted in our model, the extent of the early star loss depends very weakly on the strength of the external tidal field. As shown in Figs. \ref{fig:msgmtot1m} and \ref{fig:msgmtot4m}, the final values of \msgsp do not indeed depend significantly on the strength of the external tidal field; for the models investigated in this paper, the differences between the value of \msgsp at the end of the simulations for models in the weak field and those in the strong field range from $\sim 0$ to $\sim 0.06$.

The small effect of stellar escape due two-body relaxation on the evolution of \msgsp revealed by our simulations
could possibly play the role of a second, and much less important parameter in determining the current \msgsp of Galactic clusters. Possible observational evidence of this second-order dependence on the strength of the tidal field has been suggested in the study by Zennaro et al. (2019) and Milone et al. (2020) who found that within the main correlation between fraction of 2G stars and cluster's mass, clusters with smaller pericentric distances tend to have slightly larger 2G fractions. 

Finally we emphasize that while in this work the expansion of clusters triggered by mass loss due to stellar evolution is the process responsible for the clusters' early expansion and loss of 1G stars and evolution of \msg, there are a number of additional dynamical ingredients not included in our simulations which may further strengthen the early loss of stars. In particular, as shown in a number of studies, the complex external environment and tidal field during the first few Gyrs of clusters' evolution can lead to a significant additional loss of stars and further contribute to the evolution of \msgsp (see e.g.  Li \& Gnedin 2019, Carlberg 2020, Renaud 2018, 2020 and references therein). Primordial gas expulsion (see e.g. Boily \& Kroupa 2003) could also contribute to a cluster's early expansion and loss of stars. Finally, primordial mass segregation can strengthen the early expansion due to stellar evolution and further enhance the early rate of star escape (see e.g. Vesperini et al. 2009, Haghi et al. 2014).

\begin{figure}
	\includegraphics[width=\columnwidth]{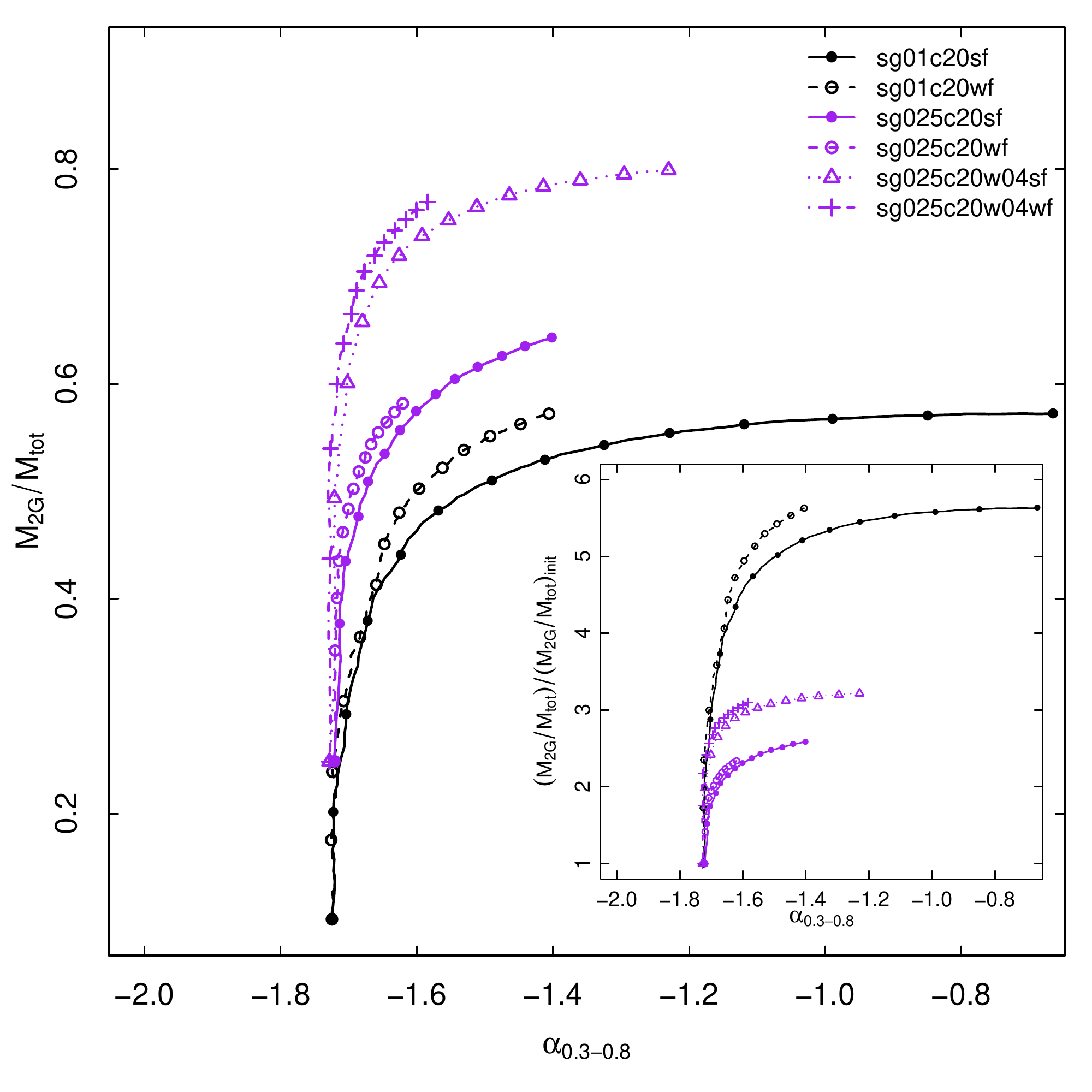}
    \caption{Evolution of the ratio of the total mass of the 2G system to the total cluster's mass, \msg, versus the power-law index of the stellar mass function for main sequence stars with masses between 0.3 \msun$~$ and 0.8 \msun, $\alpha_{0.3-0.8}$, for the following models: sg01c20sf (solid black line), sg01c20wf (dashed black line), sg025c20sf (solid purple line), sg025c20wf (dashed purple line), sg025c20w04sf (dotted purple line), sg025c20w04wf (dot-dashed purple line). Points with different symbols overplotted on each line are show the values of \msgsp and $\alpha_{0.3-0.8}$ at times between 0 Gyr and 13 Gyr with a time step of 1 Gyr. The figure in the inset shows the same data but with the values of \msgsp normalized to their initial values.}
    \label{fig:msgalpha}
\end{figure}

In order to further explore and illustrate the role of early and long-term evolutionary processes in driving the evolution of \msg, Fig. \ref{fig:msgalpha} shows for a few representative models  the evolution of \msgsp versus the power-law index of the stellar mass function, $\alpha_{0.3-0.8}$, calculated from a maximum-likelihood fit of the stellar mass function for main sequence stars with masses between 0.3 \msun$~$ and 0.8 \msun. 

As shown in several studies (see e.g. Vesperini \& Heggie 1997, Baumgardt \& Makino 2003, Trenti et al. 2010), stellar escape induced by the effects of two-body relaxation preferentially affects low-mass stars and causes a gradual flattening of the stellar mass function (hereafter MF) as a cluster evolves and loses stars. A correlation between the fraction of mass lost due to two-body relaxation and the slope of the MF has been found in several numerical studies (see e.g. Vesperini \& Heggie 1997, Trenti et al. 2010, Webb \& Leigh 2015 and references therein; see also Sollima et al. 2017, Ebrahimi et al. 2020 for observational studies). 
Assuming that the slope of the initial MF is known, the present-day MF slope can thus be used to estimate the fraction of stars lost due to two-body relaxation (Webb \& Leigh 2015, Baumgardt et al. 2019).

Early loss of stars induced by a cluster's expansion triggered by stellar evolution (or other early dynamical processes), on the other hand, does not significantly affect the cluster's stellar mass function: even the loss of a large number of stars during a cluster's early evolution phases might leave no fingerprint on the slope of the stellar MF in the low-mass range (typically 0.3-0.8 \msun) observed in old globulars clusters (see e.g. Vesperini et al. 2009).
A very extreme primordial mass segregation would be needed to affect the MF in this low-mass range with the early star loss discussed here. This aspect will be further explored in a future investigation.

Fig. \ref{fig:msgalpha} illustrates these points and their connection with the evolution of \msg: this figure  shows that all the models are characterized by an initial phase during which \msgsp increases while $\alpha_{0.3-0.8}$ is approximately constant. During this phase, the escape of a significant fraction of 1G stars leads to an increase in \msgsp but no significant evolution in $\alpha_{0.3-0.8}$. The evolutionary tracks on the $\alpha_{0.3-0.8}$-\msgsp plane clearly show that most of the loss of stars needed to increase \msgsp from its initial value to values close to those currently observed does not significantly affect the slope of the initial MF in the mass range observed in old clusters; in general, the slope of the present-day MF can not be used to infer the extent of any episode of early star loss or to rule out its role in determining the present value of \msg.
The figure in the inset shows the same data but with the values of \msgsp normalized to their initial values to further illustrate the relative variation of \msgsp during the different evolutionary phases.

Some of the models included in Fig. \ref{fig:msgalpha} lose a non-negligible number of stars due to the effects of two-body relaxation during their long-term evolution and the MF gradually flattens as a consequence of the preferential loss of low-mass stars. Examples of these systems include the sg01c20sf, sg01c20wf, sg025c20sf, and the sg025c20w04sf models. The evolution of these systems is characterized by the presence of an almost horizontal portion of the track on the  $\alpha_{0.3-0.8}$-\msgsp plane; during that phase the loss of stars driven by two-body relaxation preferentially removes low-mass stars and affects $\alpha_{0.3-0.8}$ but, as shown in Fig. \ref{fig:msgalpha} and discussed above, this loss of stars only slightly affects \msg.

\begin{figure}
	\includegraphics[width=\columnwidth]{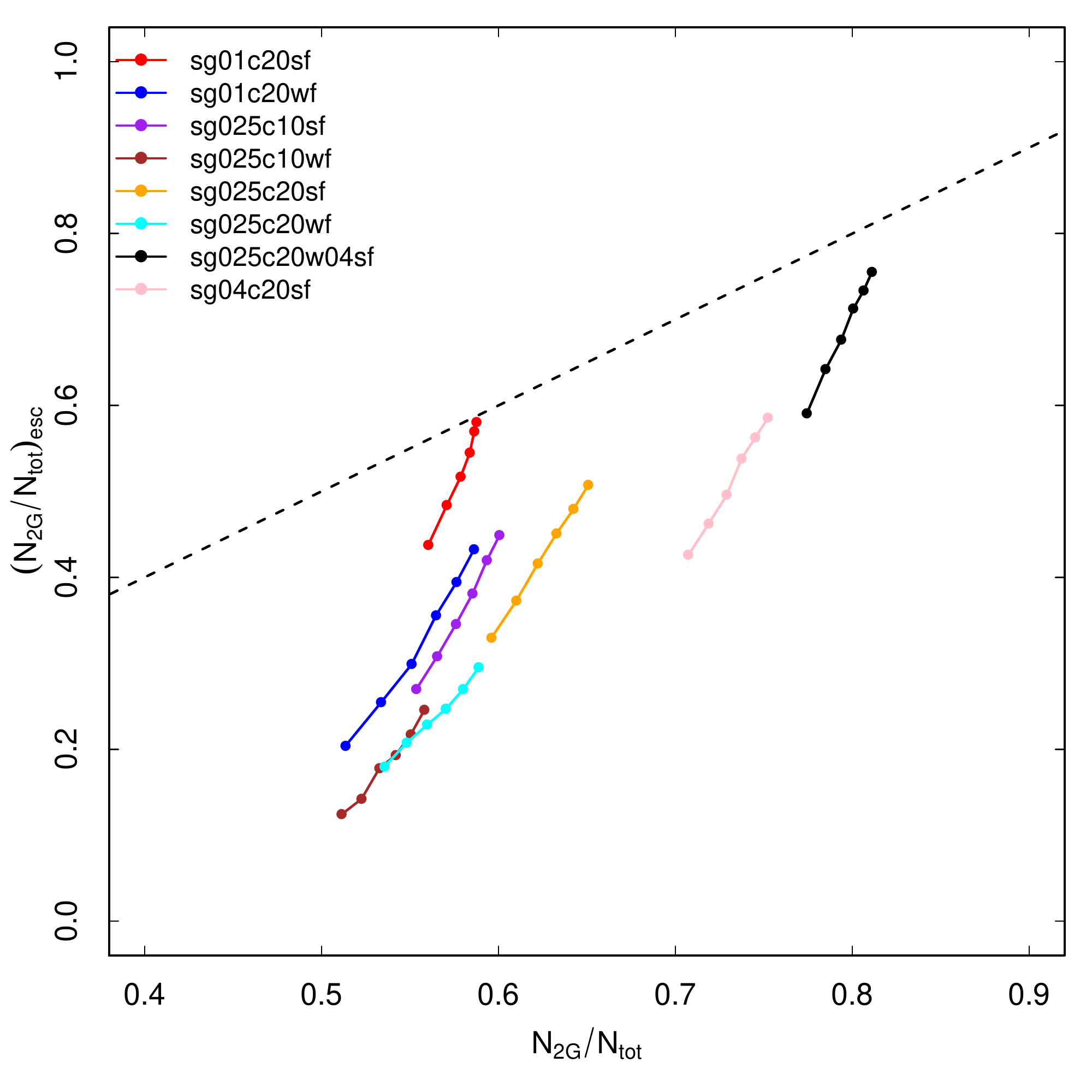}
	\includegraphics[width=\columnwidth]{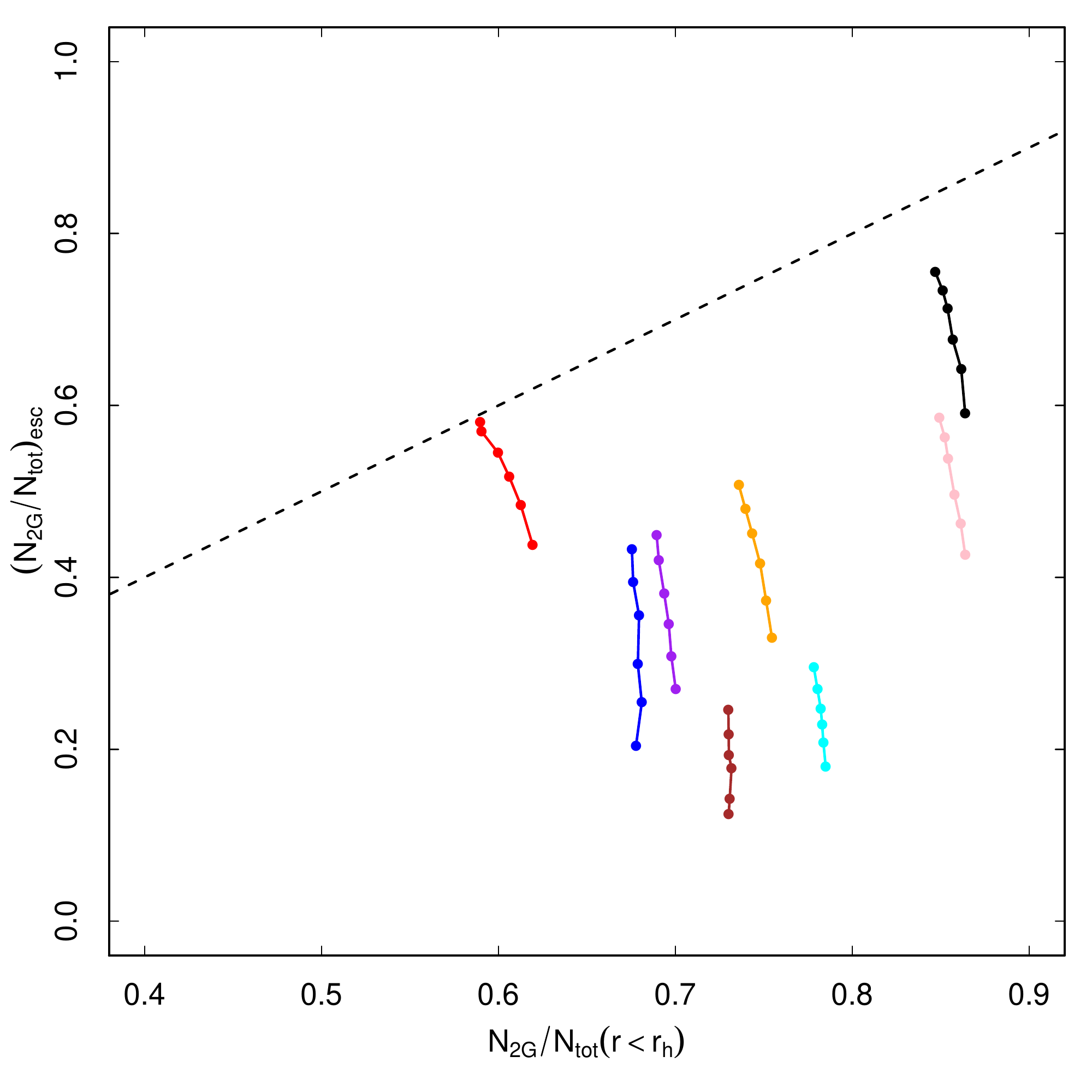}
        \caption{Evolution of the number fraction of escaping stars belonging to the 2G in time intervals of 1 Gyr between $t=8$ Gyr and $t=13$ Gyr versus the global  fraction of 2G stars inside the cluster (top panel) and versus the fraction of 2G stars inside a cluster's half-mass radius (bottom panel). For each line the point with the smallest (largest) value of escapers \escnsg$~$ corresponds to $t=8$ Gyr ($t=13$ Gyr). For each point, the value of \escnsg$~$ represents the fraction of 2G stars in the populations of stars that escaped between time $t$ Gyr and $(t-1)$ Gyr and the value of $(N_{\rm 2G}/N_{\rm tot})$ is the fraction of 2G stars inside the cluster (top panel) or inside the cluster's half-mass radius (bottom panel) at time $t$ Gyr.
Different lines correspond to the following models: sg01c20sf (red line), sg01c20wf (blue line), sg025c10sf (purple line), sg025c10wf (brown line), sg025c20sf (orange line), sg025c20wf (cyan line), sg025c20w04sf (black line), sg04c20sf (pink line). The thin dashed line represents values for which \escnsg$~$=$(N_{\rm 2G}/N_{\rm tot})$.}
    \label{fig:tidalt}
\end{figure}

We conclude this section with a brief discussion of the fraction of 2G stars among the escaping stars that populate a cluster's tidal tails. Several recent investigations have significantly extended the early observational studies (see e.g. Leon et al. 2000, Grillmair \& Dionatos 2006) of globular clusters' tidal tails (see e.g. Carballo-Bello et al. 2018, Grillmair 2019, Ibata et al. 2019, 2020, Kaderali et al. 2019, Bonaca et al. 2020a, 2020b, Thomas et al. 2020, Lane et al. 2020, Sollima 2020, Starkman et al. 2020) and revealed a variety of complex morphological and kinematic features that can shed light on the clusters' dynamics and the properties of the Galactic halo (see e.g. Sanderson et al. 2017, Bonaca \& Hogg 2018, Bonaca et al. 2019, Webb \& Bovy 2019, Reino et al. 2020). All the observational studies of multiple populations have so far focussed their attention on the clusters' more populated inner regions; future photometric and spectroscopic studies focussed on the properties of stars in the tidal tails could shed light on several key aspects of the dynamics of multiple populations although the relatively small number of stars spread in extended regions  make these studies particularly challenging (see e.g. Ji et al. 2020, Chun et al. 2020, Hansen et al. 2020, for recent studies aimed at characterizing the chemical properties of extratidal stars and stars in tidal streams possibly originating from globular clusters). Tidal tails found in observational studies are typically populated by stars escaped from a cluster in the last $\approx 0.5-2$ Gyr  (depending on the extension of the tail revealed by the observations and the cluster's orbit). Here we focus our attention on stars escaping from clusters during a more extended time interval in order to better illustrate the link with the  evolving fraction of 2G stars inside the cluster and the spatial mixing of the two populations. In Fig. \ref{fig:tidalt} we show, for a few models, \escnsg$~$ (defined as the fraction of 2G stars in the population of stars escaping from a cluster in a 1 Gyr time interval) measured at different times between 8 and 13 Gyr
versus the fraction of 2G stars inside the cluster at the same times (top panel) and the fraction of 2G stars inside the cluster's half-mass radius (bottom panel). For example, a point corresponding at $t=12$ Gyr shows on the x-axis the fraction of 2G stars inside the cluster at $t=12$ Gyr and, on the y-axis, the fraction of 2G stars in population of stars that escaped from the cluster between $t=11$ Gyr and $t=12$ Gyr.

In all the models presented here, at $t=8$ Gyr clusters have already entered the phase dominated by relaxation mass loss. As discussed above, until the two populations are spatially mixed, mass loss due to relaxation still preferentially removes 1G stars although the preferential loss of 1G stars is much weaker than that during the cluster's early evolutionary phases dominated by the expansion triggered by stellar evolution effects. The preferential loss of 1G stars implies that, in general, the fraction of 2G stars in the tidal tails is smaller than the fraction of 2G stars inside the clusters; the two fractions are identical only for clusters in which the two populations are completely mixed. Fig. \ref{fig:tidalt} shows that this is indeed the case and that the fraction of 2G stars in the population of escaping stars is in general smaller than the global 2G fraction inside the cluster. The difference between the 2G fraction in the escaping population and inside the cluster depends on the degree of spatial mixing reached by the cluster. For example, in model sg01c20sf, the two populations are completely mixed at the end of the simulation and the two fractions are indeed identical. For all the other clusters \escnsg$~$ is smaller than the fraction of 2G stars inside the cluster.

The bottom panel of Fig. \ref{fig:tidalt} shows \escnsg$~$ versus the 2G fraction inside the cluster's half-mass radius. This figure may be more useful for a comparison with future observational data for clusters that are not completely mixed since the fraction inside the cluster half-mass is often the quantity measured in observational studies.

Finally we point out that clusters that do not form 2G stars or have a small 2G fraction (see e.g. Conroy \& Spergel 2011, Dondoglio et al. 2020 for some studies that have explored the possible mass threshold for the 2G formation) could completely dissolve and leave a stream populated only or mainly by 1G stars.

\subsection{Spatial mixing and structural properties of 1G and 2G stars}
Previous studies exploring the formation of multiple populations by means of hydro simulations (see e.g. D'Ercole et al. 2008, Bekki 2010, 2011, Calura et al. 2019) have shown that 2G stars form in a compact subsystem in the central regions of the 1G system resulting in a cluster characterized by complex multi-scale structural and kinematical properties. 

\begin{figure}
	\includegraphics[width=\columnwidth]{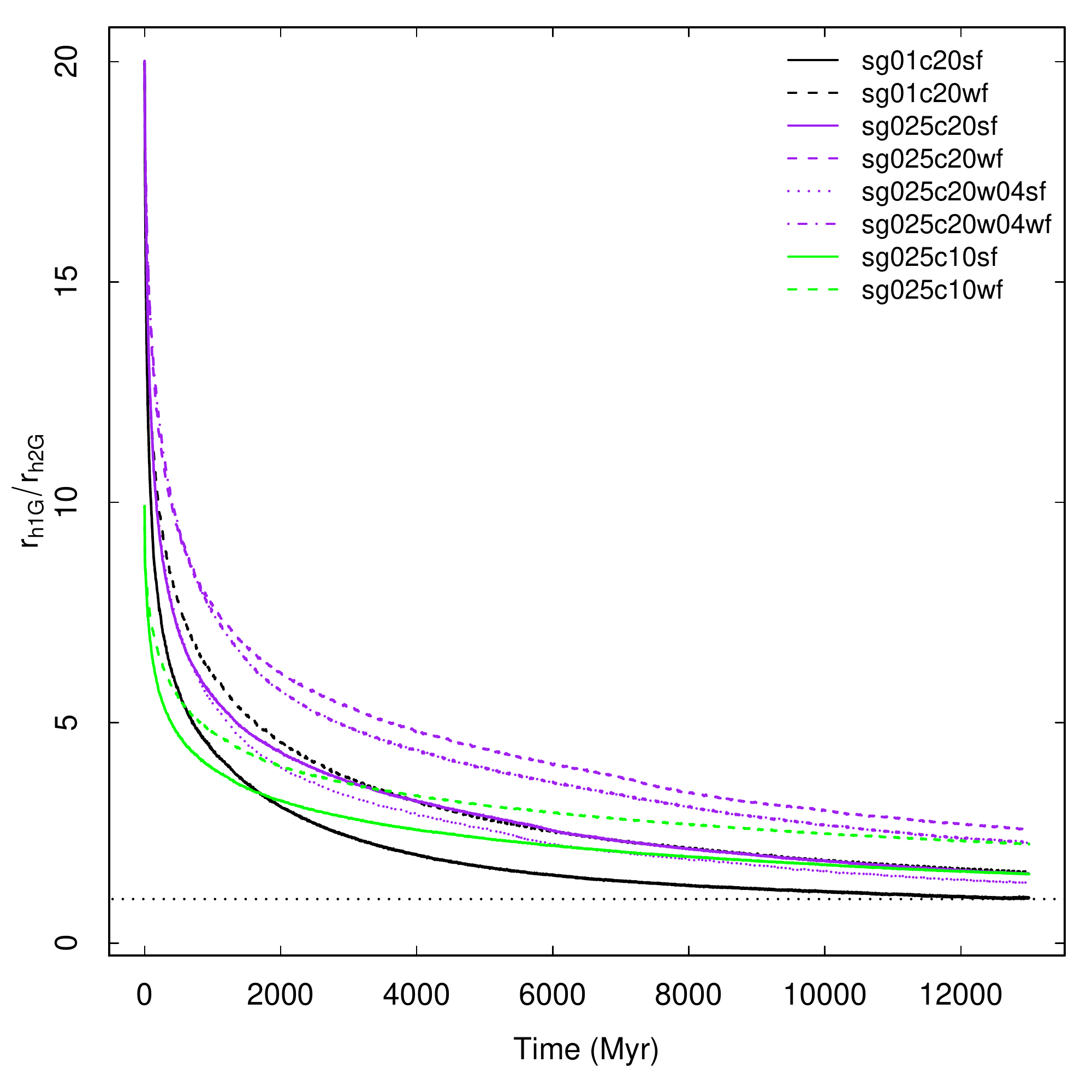}
    \caption{Time evolution of the ratio of the 1G half-mass radius to the 2G half-mass radius  for the following models: sg01c20sf (solid black line), sg01c20wf (dashed black line), sg025c20sf (solid purple line), sg025c20wf (dashed purple line), sg025c20w04sf (dotted purple line), sg025c20w04wf (dot-dashed purple line), sg025c10sf (solid green line), sg025c10wf (dashed green line). A dotted horizontal line at \rhratio$=1$ is plotted to indicate the value of this ratio corresponding to the complete spatial mixing of the two populations.}
    \label{fig:rhratiovstime}
\end{figure}

In Fig. \ref{fig:rhratiovstime} we show the time evolution of the ratio of the 1G half-mass radius to the 2G half-mass radius, \rhratio, for a few representative models. This figure clearly shows two distinct phases in the spatial mixing process. 

The first phase, occurring during the  early evolutionary stages when the cluster is losing a large fraction of its 1G population, leads to a significant decrease in \rhratio$~$ and a rapid (in a few Gyrs) evolution away from the initial structural properties set at the end of the formation process. 
For all the models studied in this paper, at the end of this first evolutionary phase \rhratio$~$ is significantly different from its initial value.  For most of the models presented here the value of \rhratio$~$ at the end of this phase falls approximately in the range $\approx 2-5$. 
 This early phase is followed by a more gradual and slower decrease of \rhratio$~$ as the cluster enters the long-term evolutionary stage in which mixing is driven by the  effects of two-body relaxation. 
At the end of the simulations, some clusters retain some differences between the spatial distributions of the 1G and the 2G spatial populations although in all cases these differences are much smaller than those in the initial conditions. Some clusters reach complete spatial mixing and, as discussed in  Vesperini et al. (2013) (see also Miholics et al. 2015, Henault-Brunet et al. 2015, Tiongco et al. 2019), complete mixing is reached when clusters  have lost a significant amount of mass during its long-term evolution (in addition to any mass loss suffered during their early evolution). This is the case, for example, for model sg01c20sf.
\begin{figure}
	\includegraphics[width=\columnwidth]{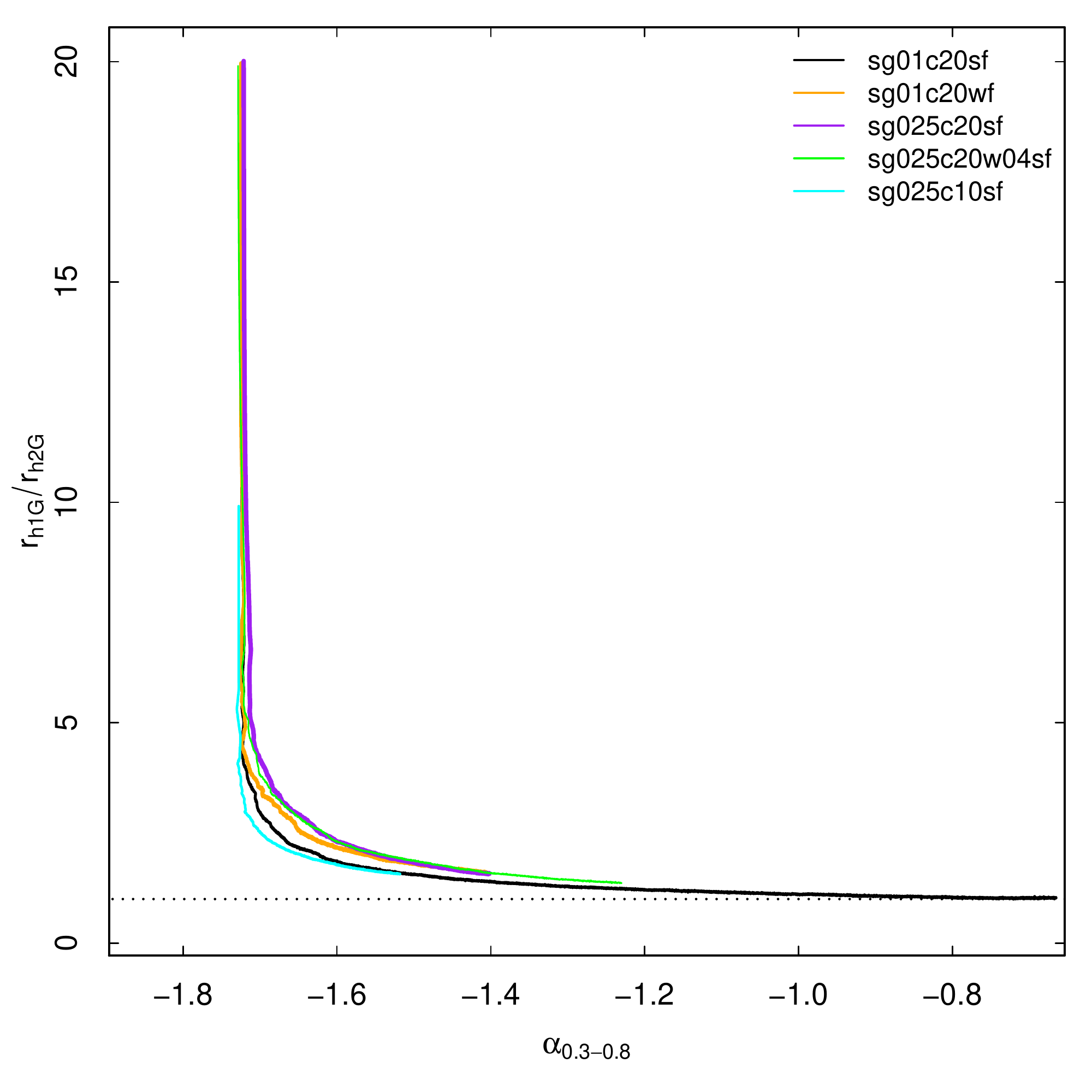}
    \caption{Evolution of the ratio of the 1G half-mass radius to the 2G half-mass radius  versus the slope of the mass function between 0.3 and 0.8 \msun, $\alpha_{0.3-0.8}$, for the following models: sg01c20sf (black line), sg01c20wf (orange line), sg025c20sf (purple line), sg025c20w04sf (green line), sg025c10sf (cyan line). A dotted horizontal line at \rhratio$=1$ is plotted to indicate the value of this ratio corresponding to the complete spatial mixing of the two populations.}
    \label{fig:rhratiovsalpha}
\end{figure}

\begin{figure}
	\includegraphics[width=\columnwidth]{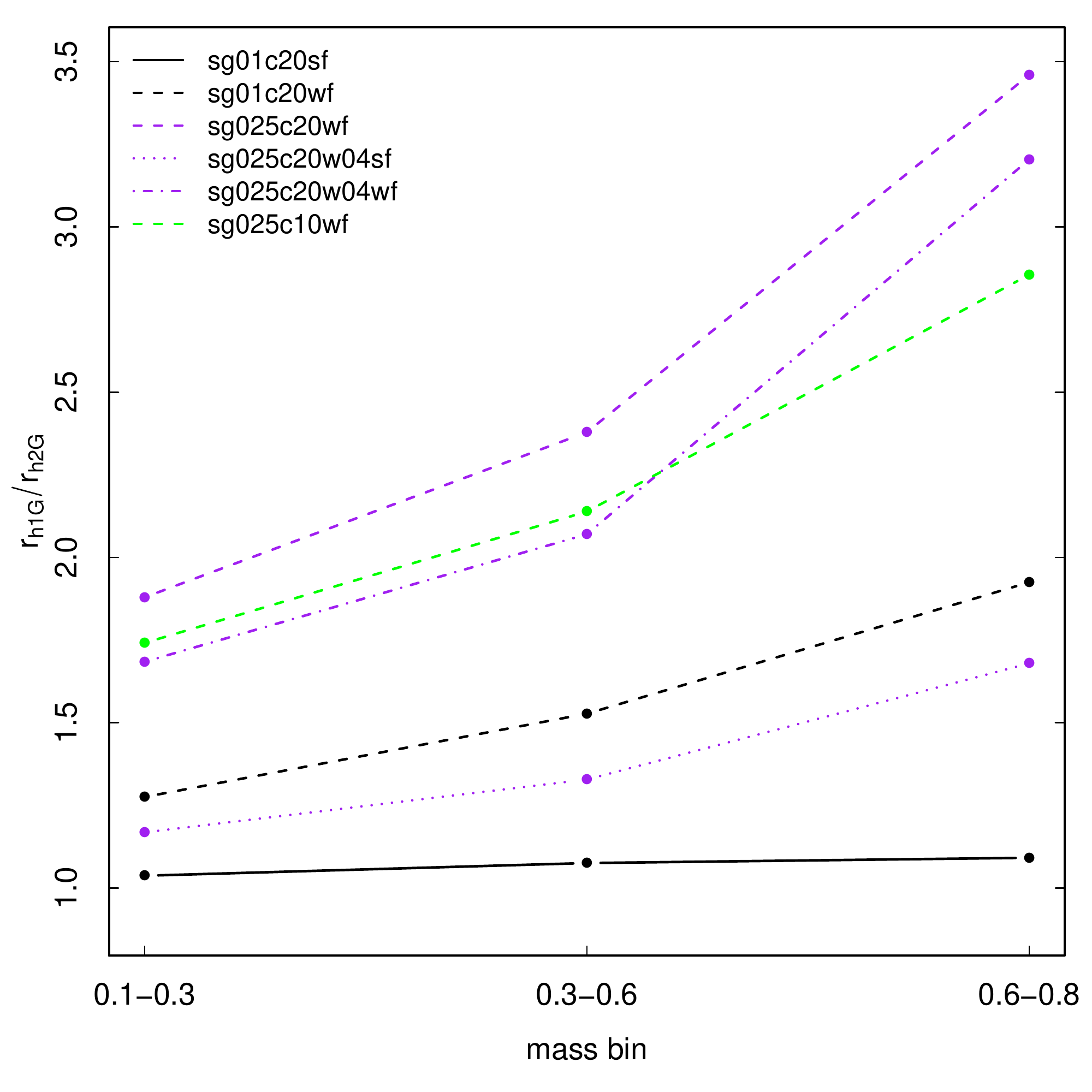}
    \caption{Ratio of the 1G half-mass radius to the 2G half-mass radius at $t=12$ Gyr for main sequence stars in three different mass bins for the following models: sg01c20sf (solid black line), sg01c20wf (dashed black line), sg025c20wf (dashed purple line), sg025c20w04sf (dotted purple line), sg025c20w04wf (dot-dashed purple line), sg025c10wf (dashed green line).}
    \label{fig:rhratiomassg}
\end{figure}
In order to further characterize these two phases in the spatial mixing process and their link to the different evolutionary stages and the mass loss occuring during a cluster's evolution, we show in Fig. \ref{fig:rhratiovsalpha} the evolution of \rhratio$~$ as a function of the slope the stellar MF, $\alpha_{0.3-0.8}$ for a few representative models. The evolution on the $\alpha_{0.3-0.8}$-\rhratio$~$ plane clearly shows the two distinct mixing phases in all models presented. During the first phase \rhratio$~$ decreases significantly while $\alpha_{0.3-0.8}$ is approximately constant; this is the early evolutionary phase during which \msgsp is rapidly increasing as a result of the preferential loss of 1G stars; as discussed in section \ref{sec:sgfraction}, the mass loss occuring  during this early phase does not affect $\alpha_{0.3-0.8}$. The second phase is instead characterized by mass loss and spatial mixing driven by the effects of two-body relaxation. The evolution of $\alpha_{0.3-0.8}$ is the consequence of the preferential loss of low-mass stars during this second phase; \rhratio$~$ continues to decrease although much more slowly.
As shown in this figure, the path followed by all clusters in the  $\alpha_{0.3-0.8}$-\rhratio$~$ plane shows little variation among the models studied; this is consistent with the results of Vesperini et al. (2013) showing that the long-term mixing of the 1G and 2G spatial distributions are closely linked with mass loss occurring during the cluster's long-term evolution.

We strongly emphasize that using the slope of the present-day MF as a general indicator of the extent of mass loss due to two-body relaxation and of the degree of spatial mixing in different clusters relies on the assumption that all clusters formed with a universal initial MF; if all clusters formed with a universal initial MF, the current value of the global MF's slope  is determined by the cluster's dynamical history and loss of stars during its long-term evolution and can be used to estimate the expected extent of spatial mixing.
On the other hand, should the stellar initial MF not be universal (see e.g. Kroupa 2020 and references therein), caution must be used in linking the expected spatial mixing with the slope of the mass function; for example, if a cluster formed with an initial MF flatter than  a standard Kroupa (2001) mass function (see e.g. Zonoozi et al. 2011, Giersz \& Heggie 2011, Webb et al. 2017, Henault-Brunet et al. 2020, Cadelano et al. 2020 for some examples of clusters for which this could be the case), a flat slope of the present-day MF does not imply that a cluster suffered  strong mass loss due to relaxation and that the different stellar populations must be completely mixed.

As already discussed in Vesperini et al. (2018), the more concentrated spatial distribution of the 2G population implies a more rapid segregation of massive 2G stars and diffusion towards the outer regions  of the low-mass 2G stars. 
One of the implications of this difference is that the timescale of 1G-2G spatial mixing depends on the stellar mass: low-mass stars tend to mix more rapidly than massive stars. Fig. \ref{fig:rhratiomassg} shows \rhratio$~$ for stars in different mass bins for a few of the models studied in this paper: with the exception of the model sg01c20sf which has reached complete spatial mixing, all the other models show that at $t=12$ Gyr \rhratio$~$ depends on the stellar mass although the trend we find is probably too weak to be detected observationally.

\begin{figure}
	\includegraphics[width=\columnwidth]{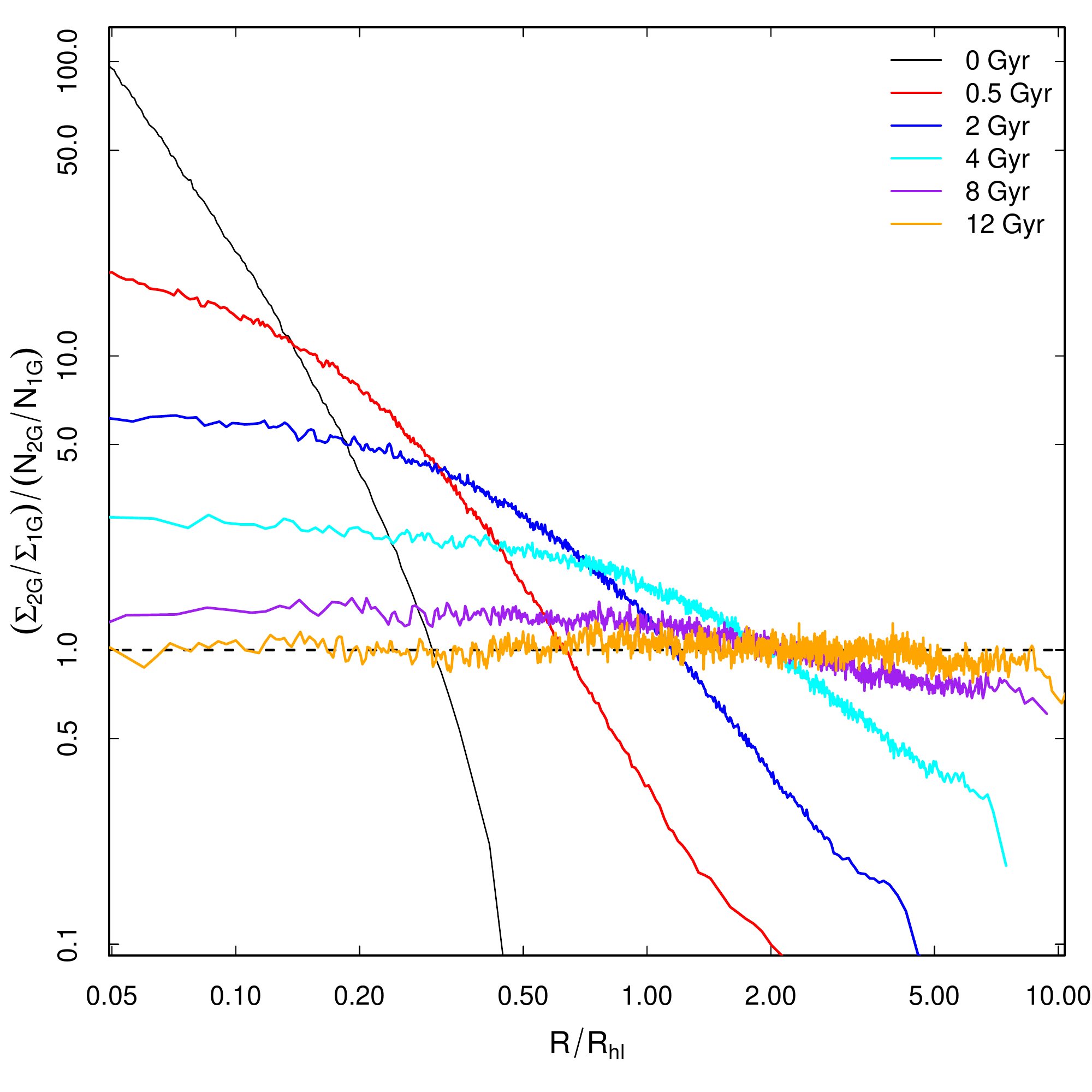}
    \caption{Radial profiles of the ratio of the 2G to the 1G surface number density profile (normalized to the global 2G to 1G number ratio) for the sg01c20sf model at t=0 (black line), 0.5 Gyr (red line), 2 Gyr (blue line) 4 Gyr (cyan line), 8 Gyr (purple line), and 12 Gyr (orange line). Radius is normalized to the cluster's half-light radius at the time the profile is calculated.}
    \label{fig:ratiosurfdens}
\end{figure}

A more detailed characterization of the internal structural properties of the two populations is provided by the time evolution of the ratio of the 2G to the 1G surface number density profiles in Fig. \ref{fig:ratiosurfdens}. We show the surface density profiles to establish a closer contact with quantities that can be determined observationally; all the profiles are calculated including only main sequence stars with masses between 0.1 and 0.8 $m_{\odot}$. The radial profile in Fig. \ref{fig:ratiosurfdens} and in all the subsequent figures show the median profile calculated from thirty different random 2D projections.  Each profile is normalized to the global 2G-to-1G number ratio so that complete mixing corresponds to a flat profile with a value equal to 1 at all radii.
This figure further illustrates the details of the dynamical path toward spatial mixing of the two populations.

\begin{figure*}
\centering{
	\includegraphics[width=5.5cm]{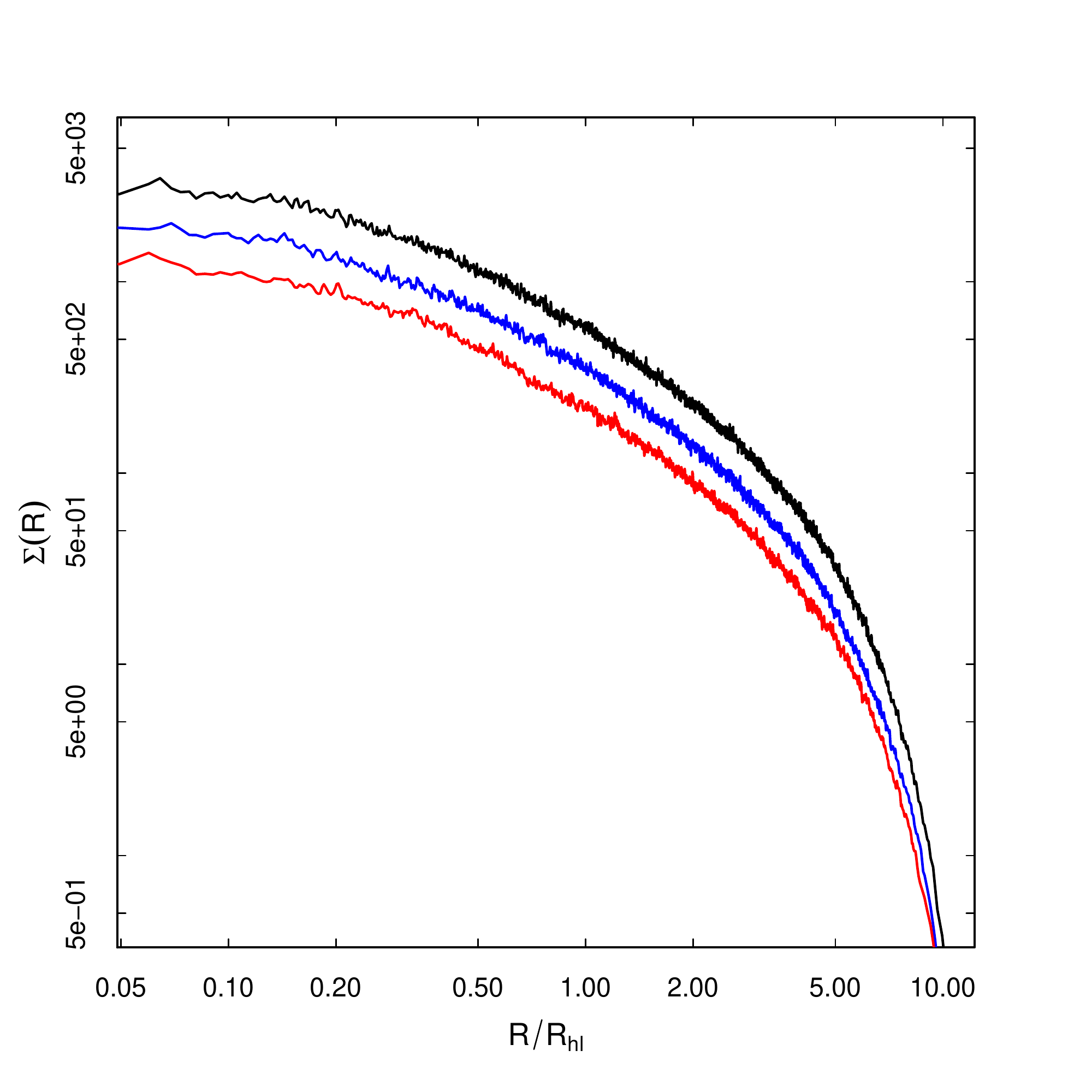}
	\includegraphics[width=5.5cm]{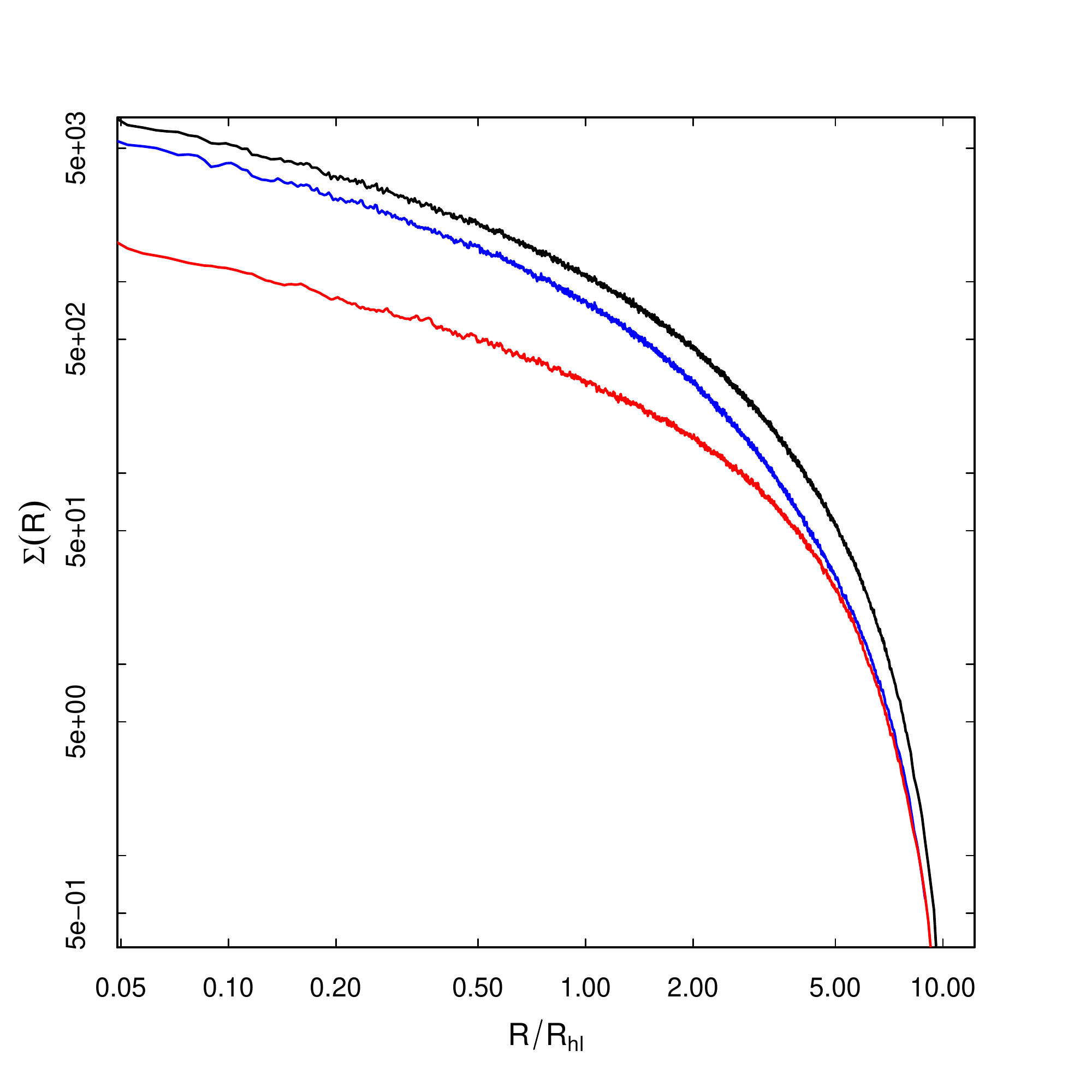}
	\includegraphics[width=5.5cm]{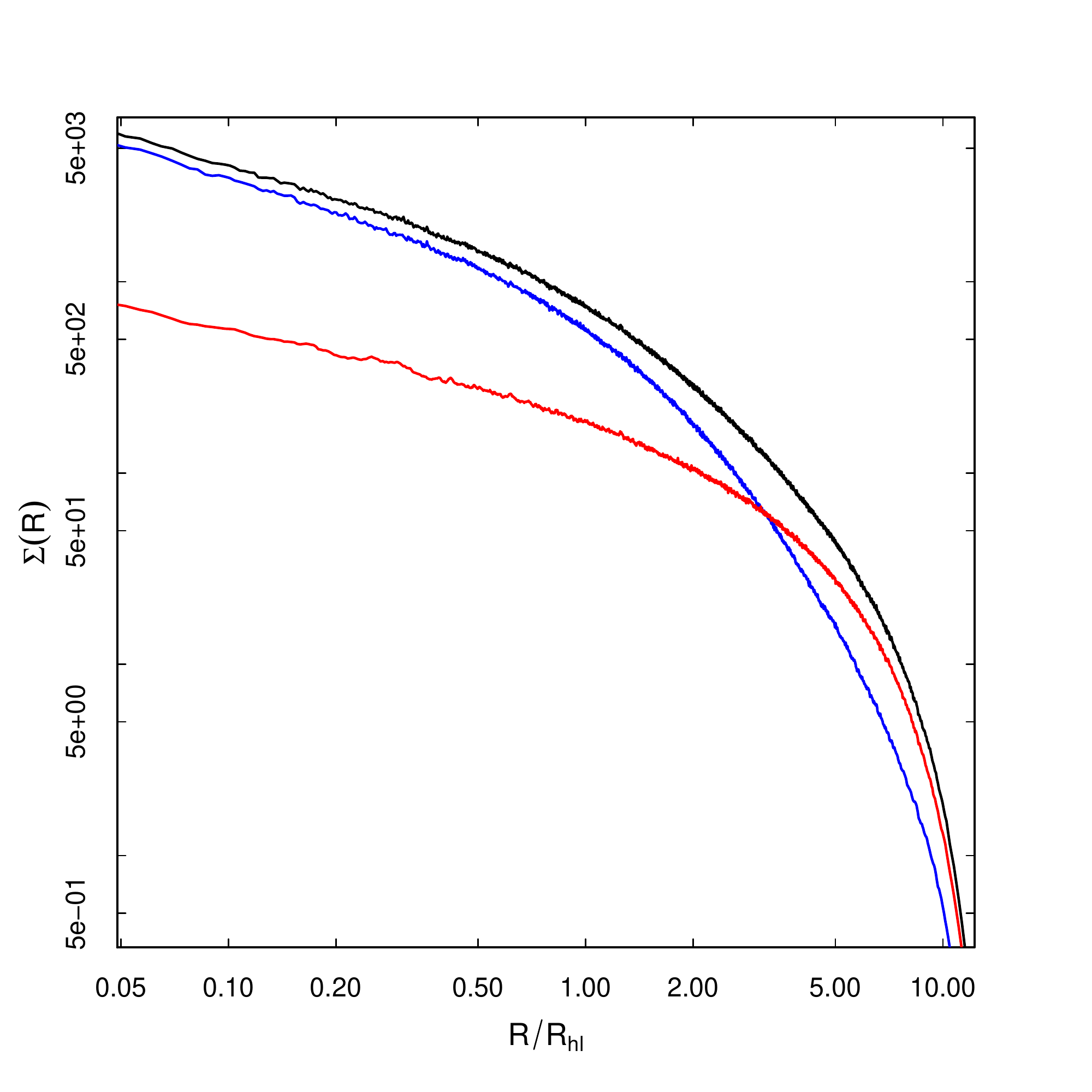}
  \caption{Surface number density radial profile of the 2G (blue line), 1G (red line) and total (black line) at $t=12$ Gyr for the sg01c20sf (left panel), sg025c20sf (middle panel), and sg025c20wf (right panel) models.}
    \label{fig:surfdens}
}
\end{figure*}

\begin{figure}
	\includegraphics[width=\columnwidth]{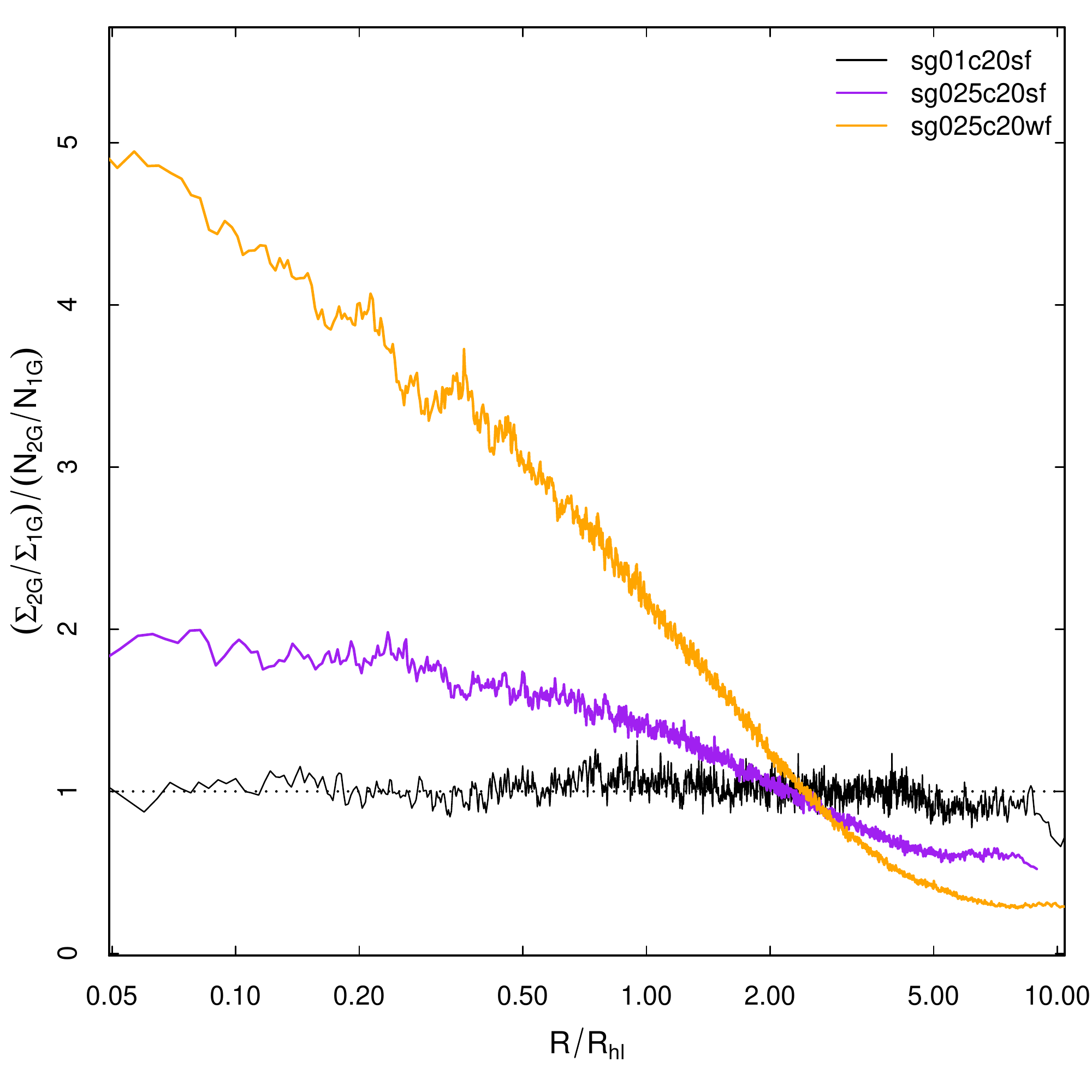}
    \caption{Radial profiles of the ratio of the 2G to the 1G surface density profile (normalized to the global 2G to 1G number ratio) at $t=12$ Gyr for the sg01c20sf (black line), sg025c20sf (purple line), and sg025c20wf (orange line) models.}
    \label{fig:ratiosurfd12}

\end{figure}

The evolution of the ratio of the 2G to the 1G density profile is driven by a combination of mixing of the two populations and the evolution of the global ratio of the number of 2G to 1G stars. 
Before complete mixing is reached the cluster 2G-to-1G surface  density ratio can be characterized by a flat portion both in the innermost and in the outermost regions: the two populations are not spatially mixed yet but their surface density profiles share a similar variation with radius in those regions leading to a flat profile of the 2G-to-1G density ratio (in some cases the outermost radial profile of the 2G profile can be slightly shallower than that of the 1G leading to a slightly increasing 2G-to-1G  surface density ratio).
Fig. \ref{fig:ratiosurfdens} also clearly shows the two phases of mixing associated with the early and long-term evolution already discussed above.

Finally in Fig. \ref{fig:surfdens} we show the final density profile for a few representative models which have reached different degrees of spatial mixing after 12 Gyr of evolution; Fig. \ref{fig:ratiosurfd12} shows the 2G-to-1G ratio of the surface density profiles at $t=12$ Gyr for the three models shown in Fig. \ref{fig:surfdens}.
In the sg01c20sf model the two populations are essentially completely mixed and follow  similar density profiles. The other two models shown in Fig. \ref{fig:surfdens}, on the other hand, are not completely mixed after 12 Gyr. In the sg025c20sf model the density profiles of the 2G and the 1G populations have a similar shape in the inner regions ($R \lesssim 0.3R_{hl}$) (which therefore corresponds to an approximately flat ratio of $\Sigma_{2G}/\Sigma_{1G}$; see the purple line in Fig. \ref{fig:ratiosurfd12}); for $R \gtrsim 0.3R_{hl}$ the $\Sigma_{2G}$ decreases more rapidly than $\Sigma_{1G}$ and finally in the outermost regions ($R \gtrsim 5R_{hl}$) the two populations follow again a similar profile (which, again, corresponds to an approximately flat ratio of $\Sigma_{2G}/\Sigma_{1G}$; see the purple line in Fig. \ref{fig:ratiosurfd12}).

The sg025c20wf model is the system that is the farthest from complete mixing among the three shown in Fig. \ref{fig:surfdens}: for this model the 2G and the 1G populations follow different density profiles  and only in the outermost regions ($R>8R_{hl}$) the two profiles have similar variation with radius resulting in a flat $\Sigma_{2G}/\Sigma_{1G}$ profile (see orange line in Fig. \ref{fig:ratiosurfd12}).

\subsection{Kinematics}
 As discussed in the Introduction, the  study of the kinematic properties of multiple stellar populations is still in its early stages but a few investigations (see e.g. Richer et al. 2013, Bellini et al. 2015, 2018, Cordero et al. 2017, Dalessandro et al. 2018, Cordoni et al. 2020a, 2020b, Lee 2020)  have started to address this aspect and further enriched our understanding of the dynamical picture of multiple populations in globular clusters.

Here we focus our attention on the evolution of the anisotropy in the velocity distribution, and the dependence of the velocity distribution on the stellar mass as the system evolves toward energy equipartition. 

\subsubsection{Anisotropy}
\begin{figure}
\centering{
	\includegraphics[width=\columnwidth]{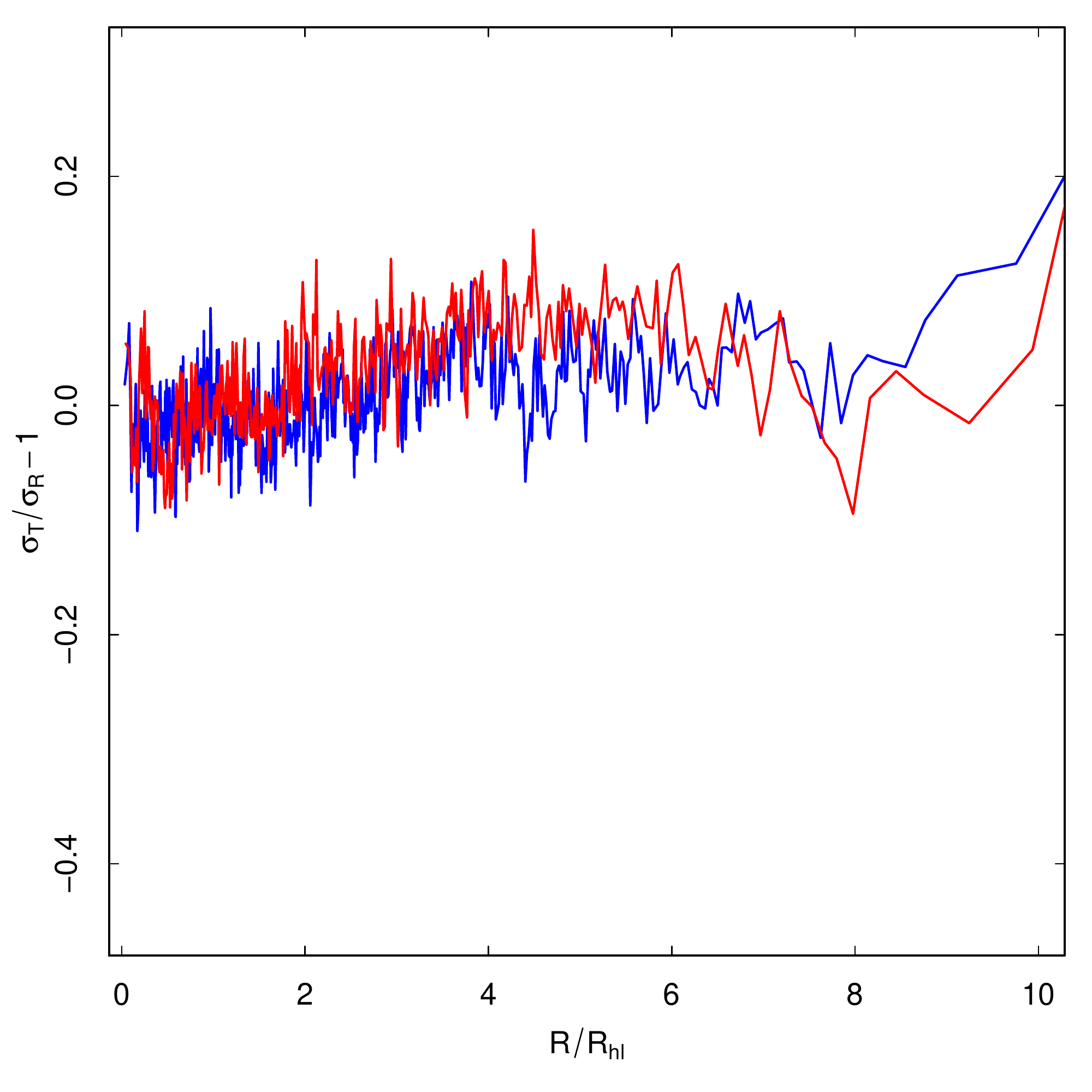}
    \caption{Radial profile of the velocity anisotropy of the 2G (blue line), 1G (red line) at $t=12$ Gyr for the sg01c20sf model.}
    \label{fig:anisotropy1}
}
\end{figure}
\begin{figure}
\centering{
	\includegraphics[width=\columnwidth]{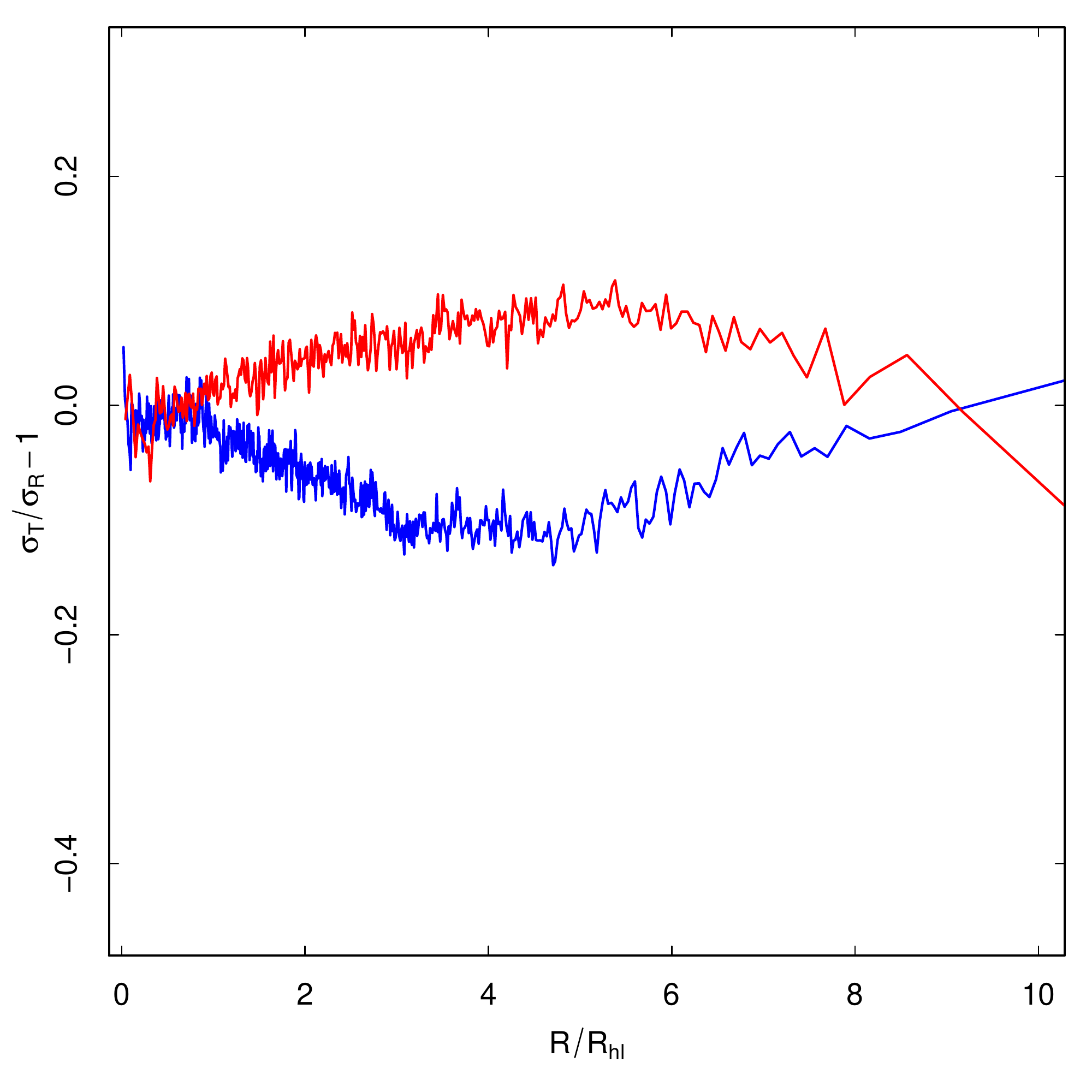}
    \caption{Radial profile of the velocity anisotropy of the 2G (blue line), 1G (red line) at $t=12$ Gyr for the sg025c20sf model.}
    \label{fig:anisotropy2}
}
\end{figure}
\begin{figure}
\centering{
	\includegraphics[width=\columnwidth]{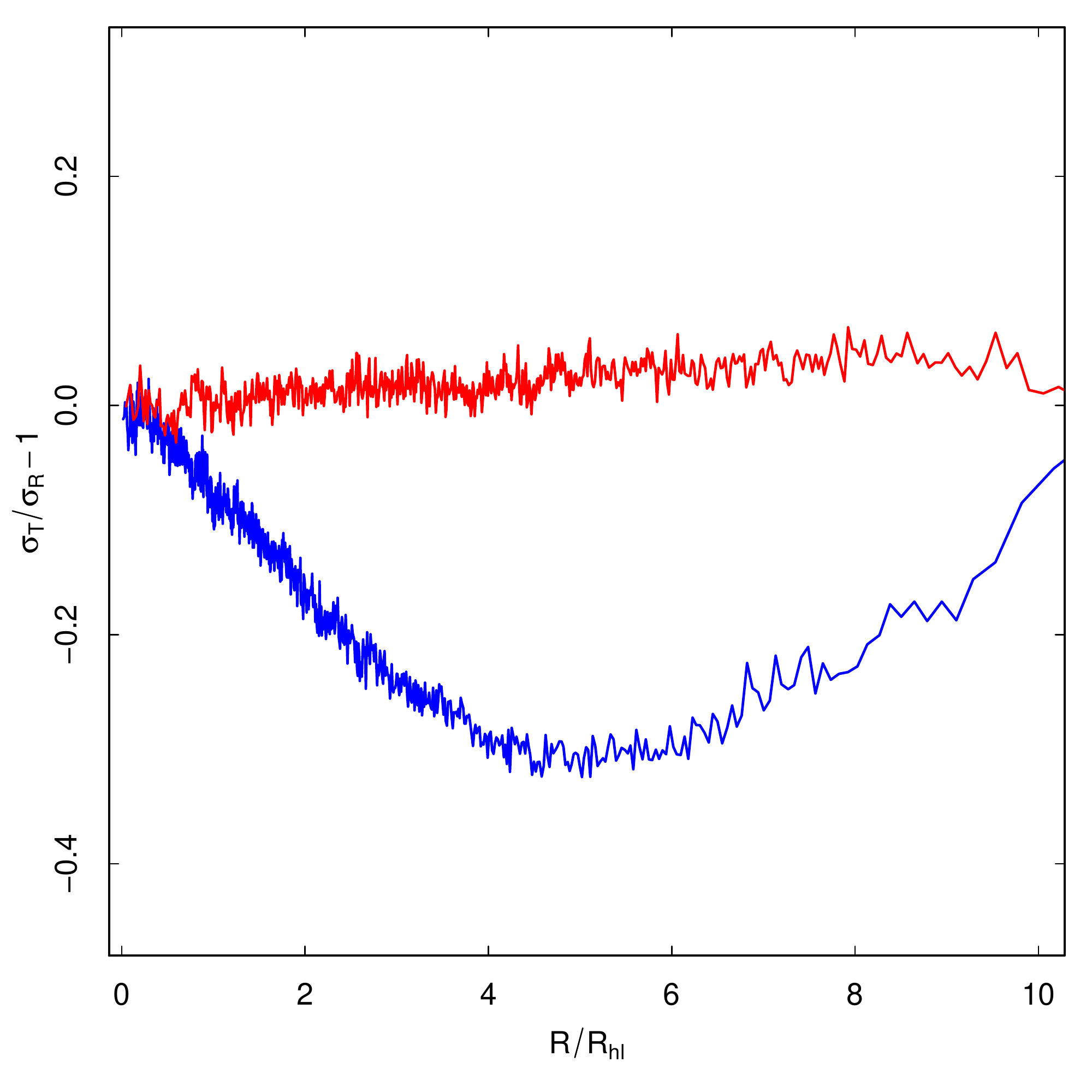}
    \caption{Radial profile of the velocity anisotropy of the 2G (blue line), 1G (red line) at $t=12$ Gyr for the sg025c20wf model.}
    \label{fig:anisotropy3}
}
\end{figure}
\begin{figure}
\centering{
  	\includegraphics[width=\columnwidth]{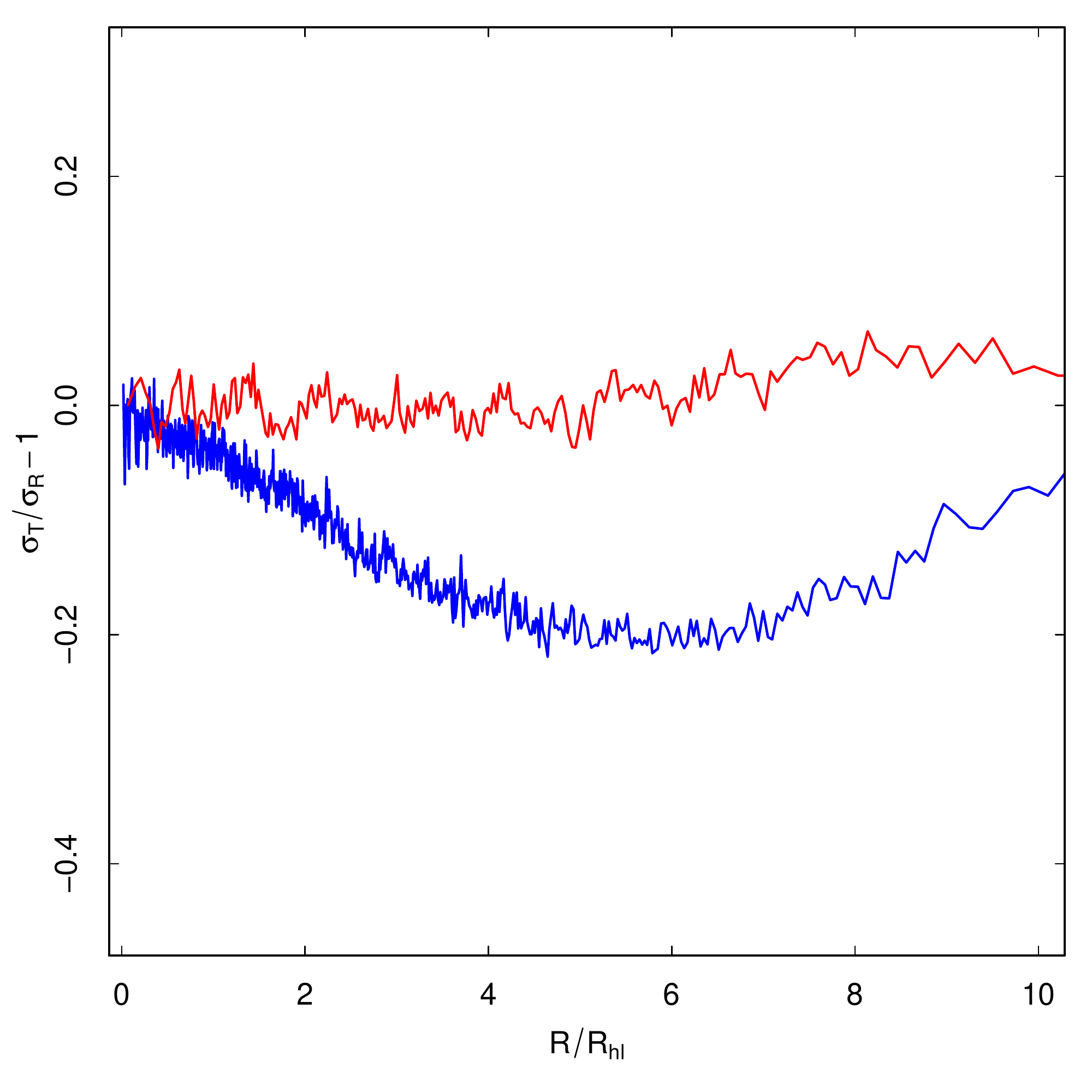}
    \caption{Radial profile of the velocity anisotropy of the 2G (blue line), 1G (red line) at $t=12$ Gyr for the model sg025c20wf with initial anisotropy radius $r_a=r_{\rm h}/2$.}
    \label{fig:anisotropyamod1}
}
\end{figure}
\begin{figure}
\centering{
  	\includegraphics[width=\columnwidth]{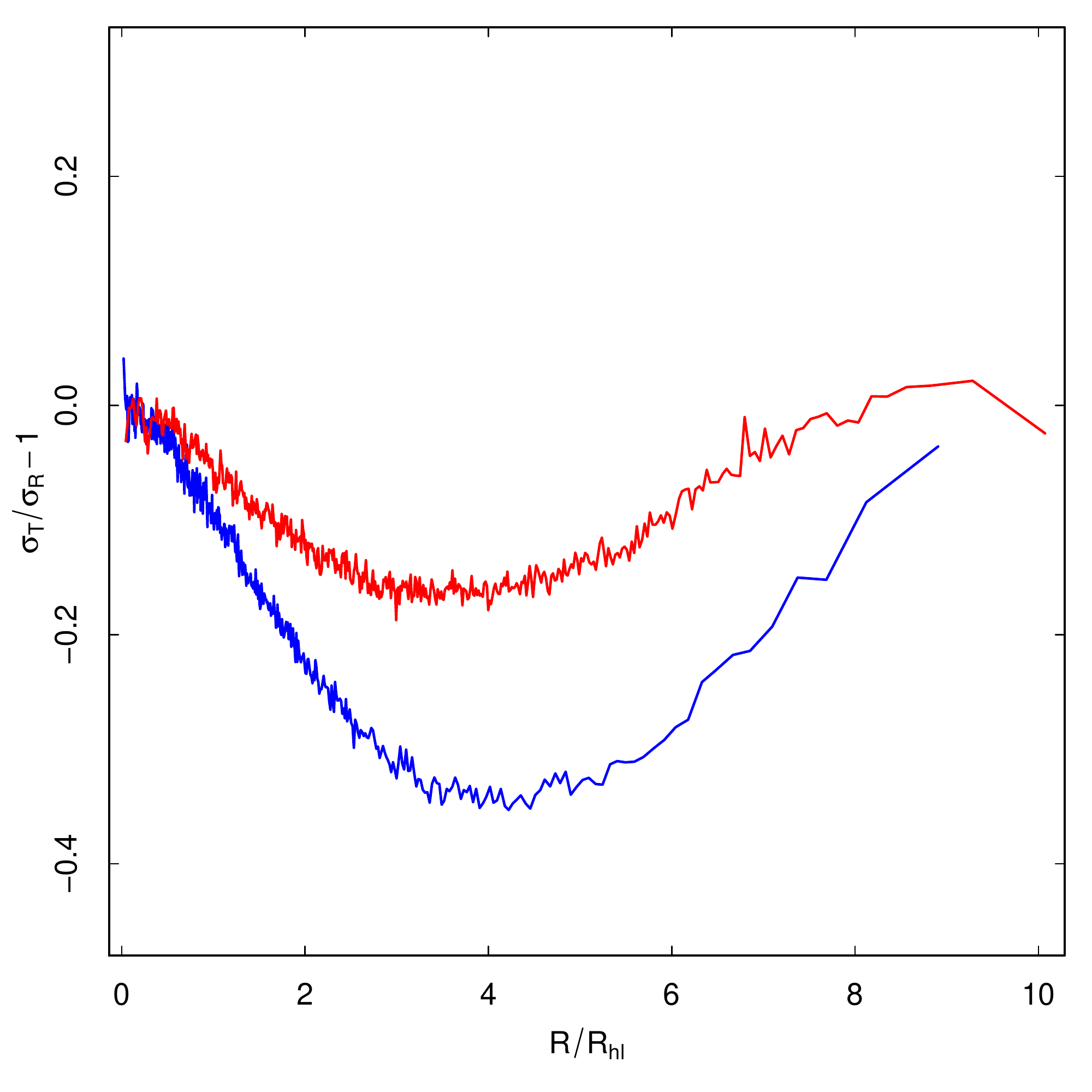}
    \caption{Radial profile of the velocity anisotropy of the 2G (blue line), 1G (red line) at $t=12$ Gyr for the model sg025c20wf with initial anisotropy radius $r_a=r_{\rm h}/3$.}
    \label{fig:anisotropyamod2}
}
\end{figure}

Figs. \ref{fig:anisotropy1}-\ref{fig:anisotropyamod2} show the projected radial profile of the anisotropy defined as \stsr, where $\sigma_T$ and $\sigma_R$ are, respectively, the radial and tangential velocity dispersion defined on the 2-D plane of projection. 

For the three models shown in Fig. \ref{fig:anisotropy1}-\ref{fig:anisotropy3}, both populations start as isotropic King models but while the 1G is initially tidally filling, the 2G is initially confined in a compact and tidally underfilling subsystem. 
These initial differences in the structural properties imply that the 2G develops a radially anisotropic velocity distribution as it diffuses from the inner to the outer regions while the 1G remains approximately isotropic (see also the discussion in Bellini et al. 2015, Tiongco et al. 2016, 2019).
 As discussed in Tiongco et al. (2016), in systems initially underfilling, the radial anisotropy developed during a cluster's evolution eventually starts to decrease as the cluster continues its evolution and can be completely erased for clusters that within a Hubble time reach the very advanced stages of their evolution and lose a large fraction of their mass. 

The three models shown in Figs.\ref{fig:anisotropy1}-\ref{fig:anisotropy3}  show the anisotropy profiles of three clusters which have reached different stages of their dynamical evolution at $t=12$ Gyr: the sg01c20sf (Fig. \ref{fig:anisotropy1}) is the model in the most advanced dynamical phase (as illustrated also by the fact that at $t=12$ Gyr the two populations are spatially mixed; see Fig. \ref{fig:ratiosurfd12}): the radial anisotropy developed by the 2G during its evolution has been erased by the effects of relaxation and mass loss and both populations have a now similar anisotropy profile (isotropic in the inner regions and slightly tangentially anisotropic in the outer regions).

The other two models shown in Fig. \ref{fig:anisotropy2} and  Fig. \ref{fig:anisotropy3} are in a less advanced dynamical phase and in both of them the 2G shows the signature of the anisotropy radial profile developed during its previous evolution and diffusion towards the cluster's outer regions: the 2G is isotropic in the innermost regions ($R<R_{hl}$) and radially anisotropic in the intermediate/outer regions. In the outermost regions, the radial anisotropy of the 2G population decreases again and the velocity distribution turns into an approximately isotropic/tangentially anistropic distribution.

It is  important to emphasize that while in these models the radial anisotropy develops during the cluster's long-term evolution, it could also develop for both the 1G and the 2G populations during the early formation and violent relaxation phases of the cluster evolution (see e.g. Vesperini et al. 2014, Tiongco et al. 2016).

In Fig. \ref{fig:anisotropyamod1} and Fig. \ref{fig:anisotropyamod2} we show the anisotropy profile for two systems starting with initial  conditions characterized by a radially anisotropic velocity distribution. The two models start with different degrees of initial anisotropy (one with anisotropy radius, $r_a$, equal to $r_{\rm h}/2$ and the other equal to $r_{\rm h}/3$; see section  \ref{sec:methods}) and reach different stages of their dynamical evolution.
For the model with $r_a=r_{\rm h}/2$  (Fig. \ref{fig:anisotropyamod1}) mass loss and dynamical evolution have erased the initial anisotropy of the 1G population and the 1G is approximately isotropic while the 2G is isotropic in the inner and outermost regions and radially anisotropic in the intermediate regions. In the model $r_a=r_{\rm h}/3$ (Fig. \ref{fig:anisotropyamod2}), on the other hand, the effects of dynamics and mass loss are not sufficient to completely erase the initial anisotropy of the 1G population. Also in this case the 2G is more radially anisotropic than the 1G but the 1G itself is characterized by a moderate radial anisotropy.

\begin{figure}
\centering{
	\includegraphics[width=\columnwidth]{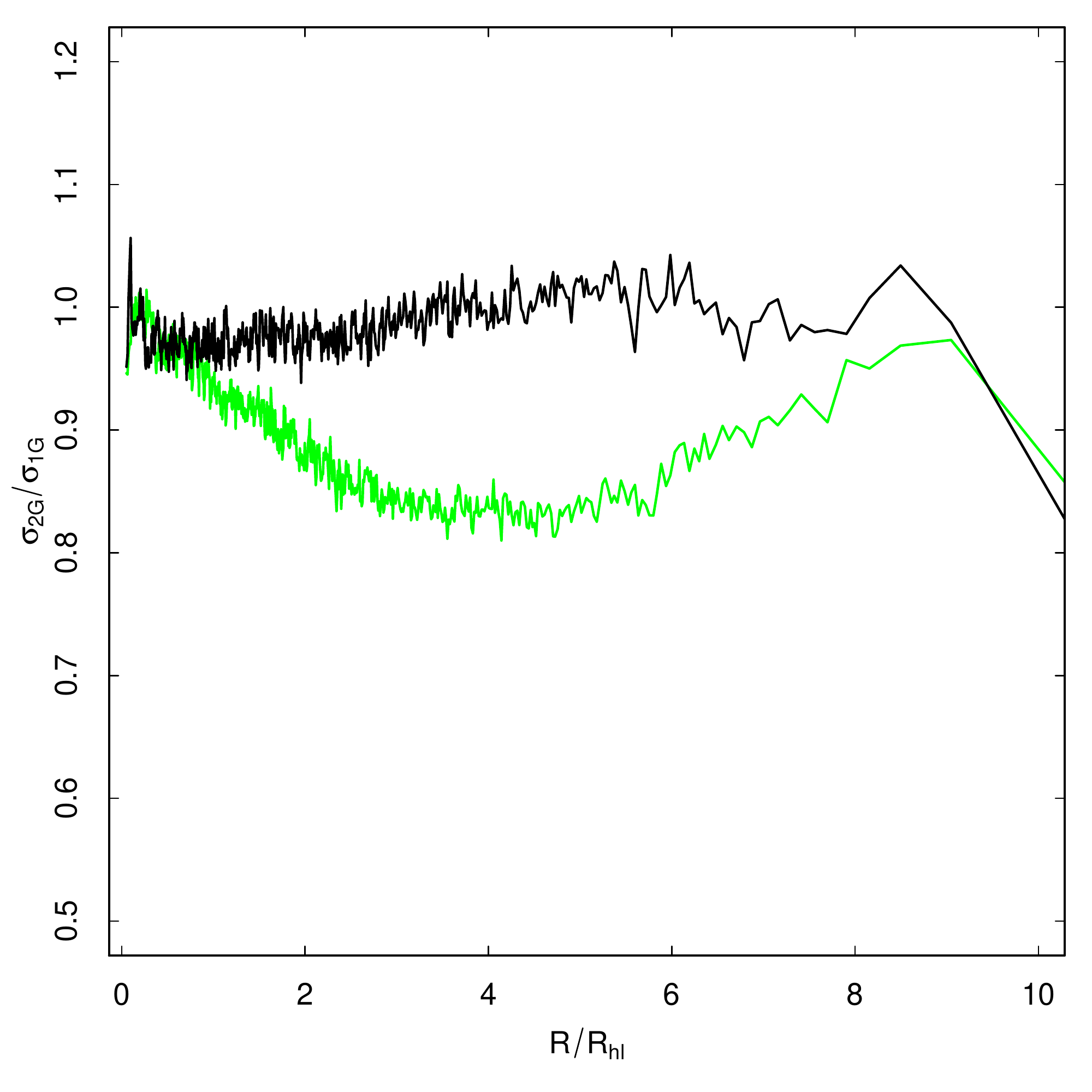}
	\includegraphics[width=\columnwidth]{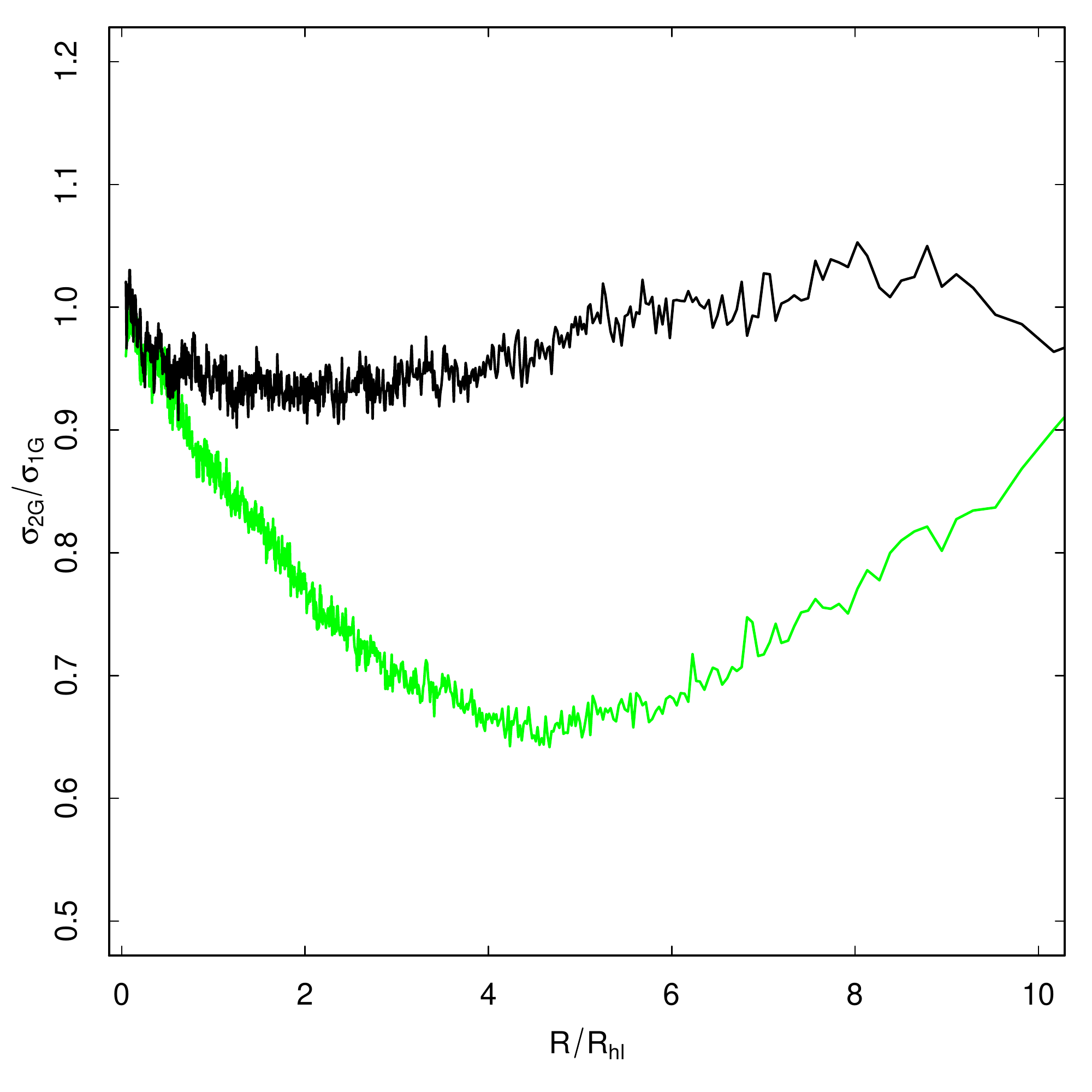}
    \caption{Radial profile of the ratio of the 2G to the 1G $\sigma_{R}$ (black line) and of the ratio of the 2G to the 1G $\sigma_T$ (green line) at $t=12$ Gyr for the sg025c20sf (top panel), and sg025c20wf (bottom panel) models.}
    \label{fig:vdispratio}
}
\end{figure}

Finally, in Fig. \ref{fig:vdispratio} we show the projected radial profile of the ratio of the tangential velocity dispersions of the 1G and the 2G populations and of the 2G to the 1G radial velocity dispersions at $t=12$ Gyr for the models sg025c20sf and sg025c20wf (see Figs.\ref{fig:anisotropy2} and \ref{fig:anisotropy3} for the anisotropy profiles of these models). This figure shows that the differences between the 1G and 2G  radial anisotropy is due mainly to the smaller tangential velocity dispersion of the 2G population. This is in agreement with what found in a few observational studies (Richer et al. 2013, Bellini et al. 2015, Milone et al. 2018, Cordoni et al. 2020a; see also Bellini et al. 2015 and Tiongco et al. 2019 for N-body simulations showing the same trend).

We will present in a future study a more comprehensive investigation of the evolution of clusters with different initial kinematic properties; the results presented here show that the models considered in this study and characterized by a tidally filling 1G and a compact 2G population always result in a 2G more radially anisotropic than the 1G population (or with both populations characterized by an isotropic distribution if the cluster is its advanced dynamical stages). The 1G population can be either isotropic or radially anisotropic  depending on the cluster's early and long-term dynamical history.
It is interesting to notice that the 1G and 2G anisotropy profiles in some of our models approximately follow those observed in the few clusters in which the kinematics of multiple populations has been studied (see e.g. Richer et al. 2013, Bellini et al. 2015, Milone et al. 2018, Cordoni et al. 2020a, 2020b): in some clusters both populations are approximately isotropic or slightly tangentially anisotropic while in other clusters the 1G is approximately isotropic at all radii explored while the 2G is radially anisotropic with an anisotropy varying with the distance from the cluster's centre as found in our simulations.

\subsubsection{Energy equipartition}
\begin{figure}
\centering{
  	\includegraphics[width=\columnwidth]{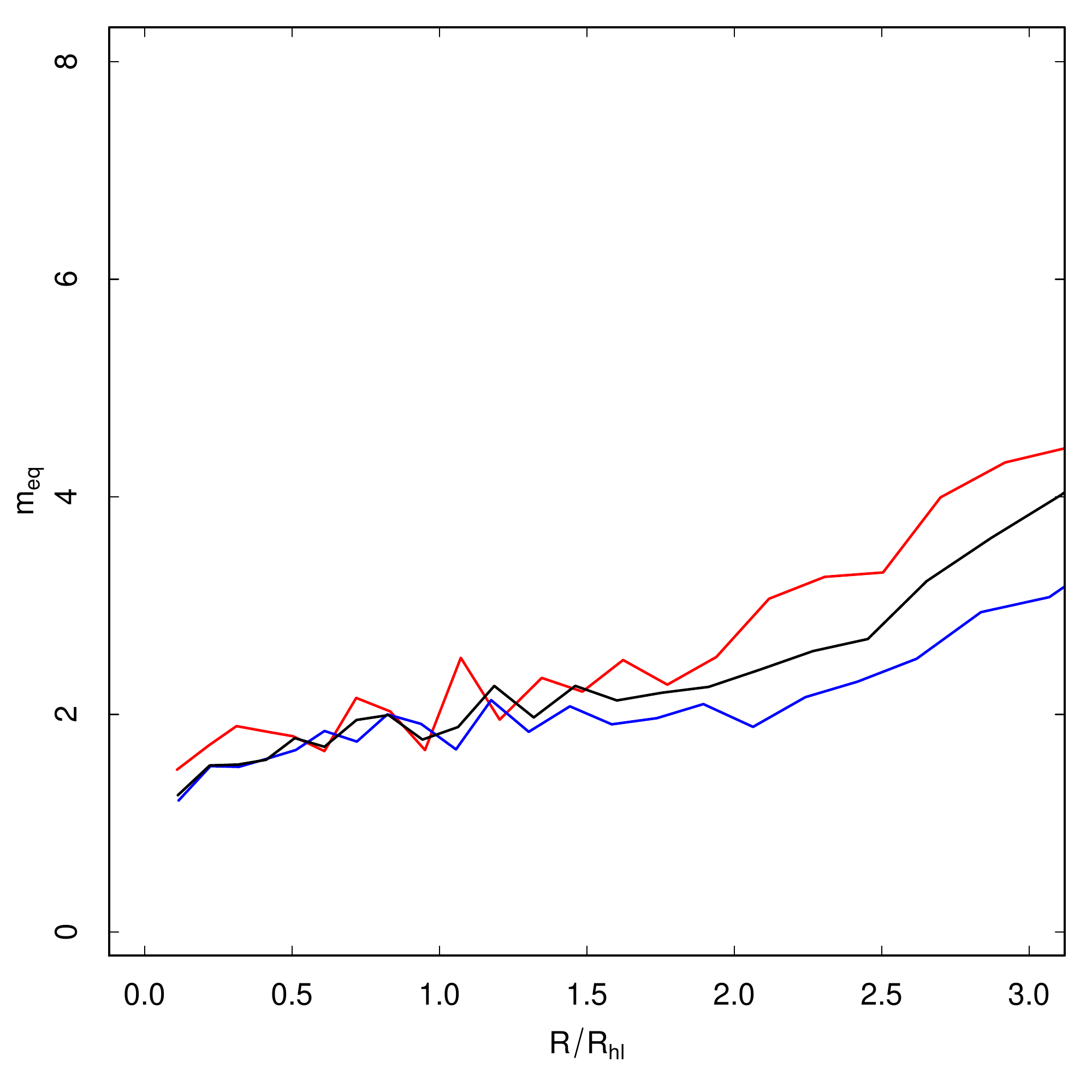}
  	\includegraphics[width=\columnwidth]{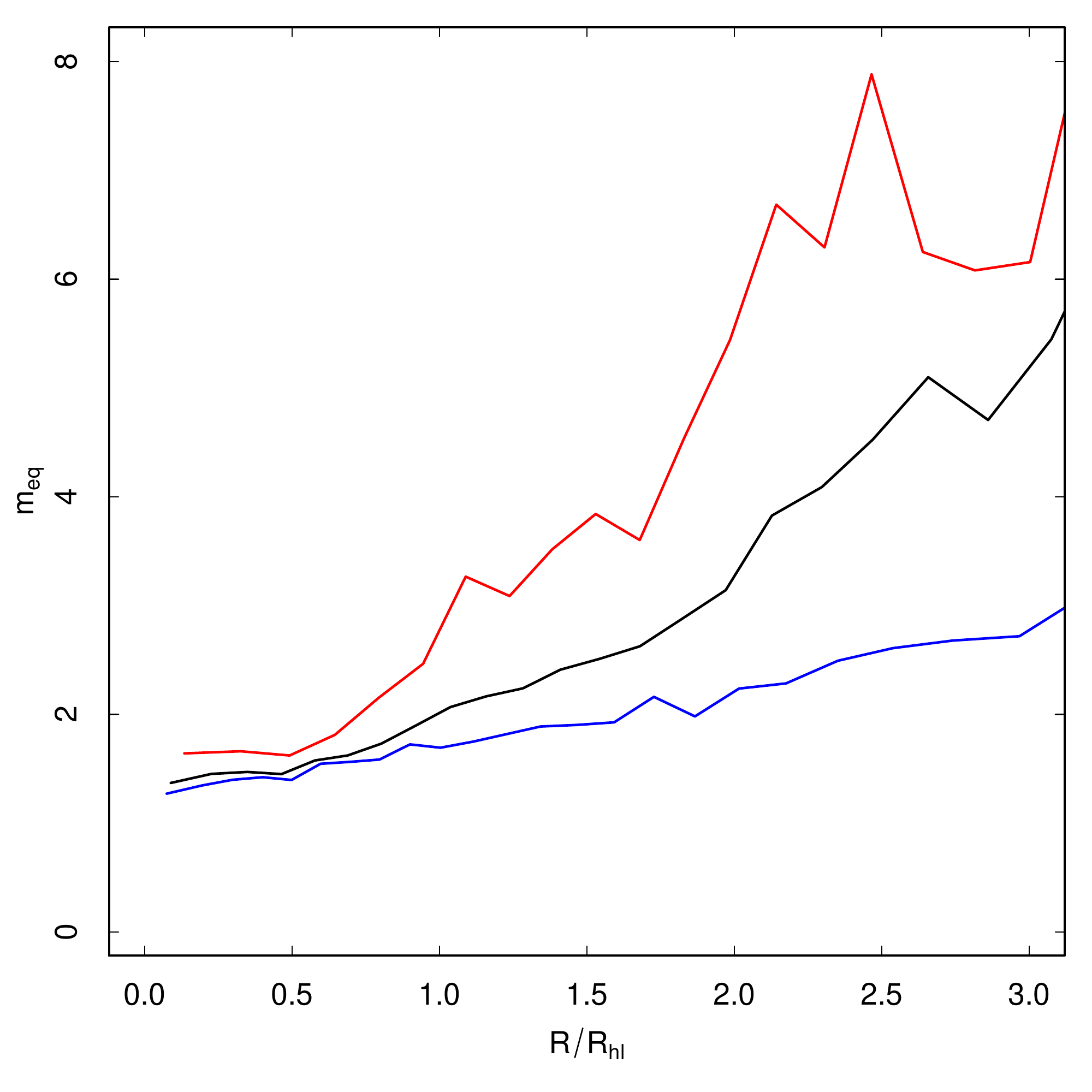}
        \caption{ Radial profile (with radius normalized to the half-ligh radius) of the equipartition mass, $m_{\rm eq}$, (see section 3.3.2) for the sg01c20sf model (top panel) and the sg025c20sf model (lower panel) at $t=12$ Gyr. In each panel the values of $m_{\rm eq}$ for the 1G (red line), the 2G (blue line) and the 1G and the 2G combined (black line) are shown.}
    \label{fig:equipmeq}
}
\end{figure}

The initial differences in the structural properties of multiple populations have dynamical implications for the evolution of the 1G and 2G populations toward energy equipartition. In order to measure the degree of energy equipartition reached by a given population, we use the exponential function  introduced by Bianchini et al. (2016) to fit the variation of the velocity dispersion, $\sigma(m)$, with the stellar mass $m$, $\sigma(m) \propto exp(-0.5 m/m_{eq})$. Smaller values of $m_{eq}$ correspond to higher degree of energy equipartition (see Bianchini et al. 2016 for further discussion). For our analysis we have used to total velocity dispersion measured on a 2-D plane of projection and focussed our attention on main sequence  stars with masses between 0.1 and 0.8 $m_{\odot}$. Fig. \ref{fig:equipmeq} shows the radial variation of $m_{\rm eq}$ at $t=12$ Gyr for two of the models studied in this paper. As expected in both cases the inner regions are characterized by smaller values of $m_{\rm eq}$ and, therefore, by a higher degree of energy equipartition. This is the consequence of the fact that the inner regions are those characterized by shorter relaxation times. The two models, however, show different behavior in the radial variation of $m_{\rm eq}$ for the 1G and the 2G populations. As discussed in the previous sections, the system sg01c20sf is in its advanced evolutionary stages: at $t=12$ Gyr the two populations are spatially mixed and characterized by an isotropic velocity distribution. For this model the degree of energy equipartition of the two populations is very similar for $R\lesssim 2R_{\rm hl}$ while in the outer parts of the radial range explored the 2G is characterized by slightly smaller values of $m_{\rm eq}$.

Model sg025c20sf, on the other hand, is in a less advanced evolutionary stage, its populations are not completely mixed (see Fig. \ref{fig:ratiosurfd12}) and are characterized by velocity distributions with different levels of anisotropy (Fig. \ref{fig:anisotropy2}). For this system (see lower panel of Fig. \ref{fig:equipmeq}), outside the cluster's innermost regions ($R\gtrsim 0.5 R_{\rm hl}$) the values of $m_{\rm eq}$ for the 2G population are significantly smaller than those of the 1G population clearly showing that the initially more compact 2G system is in a more advanced phase of its evolution toward energy equipartition than the 1G system. In this case the difference between the values of $m_{\rm eq}$ of the 1G population and the 2G population provides an interesting signature of the initial structural differences of the two populations.  The values of $m_{\rm eq}$ calculated for the 1G and 2G stars together are closer to those of the 2G in the inner regions where 2G stars are the dominant population (since the two populations are not mixed yet) and gradually become more similar to those of the 1G population in the outer regions which are predominantly populated by 1G stars.

\section{Conclusions}
In this paper, we have presented the results of a set of Monte Carlo simulations following the dynamical evolution of multiple-population clusters starting with different initial cluster masses,  2G mass fractions and density profiles, 2G-to-1G size ratios, and tidal radii.

We explored the role of early and long-term dynamical processes in driving the evolution of the main internal structural and kinematic properties of multiple populations, illustrated how the initial differences in the dynamical properties of 1G and 2G stars evolve and how, in turn, they may lead the 1G and 2G populations on different evolutionary paths.

The main results of this study are the following.

\begin{itemize}
\item In our initial conditions, we have explored systems with an initial fraction of the total mass in 2G stars, \msg, ranging from 0.1 to 0.4.  In order for \msgsp to reach values in the range of those observed ($\sim 0.35-0.9$), clusters need to preferentially lose 1G stars. In our simulations most of the evolution of \msgsp occurs during the first few Gyrs of a cluster's evolution when the system is responding to the mass loss due to stellar evolution by expanding and preferentially losing 1G stars which are less centrally concentrated than the 2G stars. At the end of this early phase, \msgsp has increased and reached values within the observed range (see Figs. \ref{fig:msgmtot1m}-\ref{fig:msgmtot4m}). We show that this early loss of stars does not affect the slope of the stellar mass function (see Fig. \ref{fig:msgalpha}) and, therefore, mass loss estimates based on the slope of the present-day mass function can not be used to infer the extent of this early episode of star loss. Additional early dynamical processes not included in our analysis (e.g. tidal shocks, expansion triggered by primordial gas expulsion) could further increase the number of 1G stars escaping during this early evolutionary phases and contribute to the evolution of \msg.
\item Once a cluster enters the long-term evolution phase dominated by the effects of two-body relaxation, the relaxation-driven loss of stars  causes only a slight increase in \msgsp since in this phase clusters, while still preferentially losing 1G stars, start to lose also 2G stars (see Figs. \ref{fig:msgmtot1m}-\ref{fig:msgalpha}). 
\item The final values of \msgsp for our models span a range between $0.53$ and $0.8$ and the final total number of stars ranges from $\sim 5\times 10^4$ to $\sim 1.7\times10^6$ (see Fig. \ref{fig:msgvsnfni} and Table 2).
\item The fraction of 2G stars in the population of escaping stars, \escnsg, varies during a cluster's evolution. 
  During the early phases of a cluster's evolution the population of escapers is dominated by 1G stars. Later in the cluster's evolution \escnsg$~$ falls in the range 0.3-0.7 depending on the fraction of 2G stars in the cluster as well as on the degree of spatial mixing of the two populations.
  In all cases \escnsg$~$ is smaller than or equal to the fraction of 2G stars inside the cluster (see Fig. \ref{fig:tidalt}).
  Clusters that do not form 2G stars or form only a small 2G population could completely dissolve and leave a stream populated only or mainly by 1G stars.
\item As a cluster evolves, the 1G and 2G populations mix and the differences between the spatial distributions of the two populations decrease. We have identified two distinct phases in the spatial mixing process: an early phase when the ratio of the 1G to the 2G half-mass radii, \rhratio, rapidly decreases followed by a more extended phase characterized by a slower mixing (see Figs. \ref{fig:rhratiovstime}-\ref{fig:rhratiovsalpha}). We confirm the results of our previous study and find that complete mixing is reached after the cluster has lost a significant fraction of mass during its long-term evolution.We find that low-mass stars mix more rapidly than more massive stars although the expected differences between \rhratio$~$  for low-mass stars and more massive stars are small (see Fig. \ref{fig:rhratiomassg}).
  For a few representative models we have presented the complete final surface density radial profile and shown in more detail the differences and similarities of the 1G and 2G density profiles (see Figs. \ref{fig:ratiosurfdens}-\ref{fig:ratiosurfd12}).
\item We have studied the evolution of the cluster's internal kinematics and found that the 2G population is characterized by a radially anisotropic velocity distribution. The extent of the 2G radial anisotropy varies with the distance from the cluster's centre: the velocity distribution is isotropic in the innermost regions, it becomes increasingly anisotropic at larger distances from the centre and finally becomes isotropic or slightly tangentially anisotropic again in the outermost regions. The 1G population can either be isotropic at all clustercentric distances or be characterized  by radial anisotropy profile similar to that of the 2G but with radial anisotropy which is in all cases weaker than that of the 2G population.
  The differences between the 1G and the 2G radial anisotropy are due mainly to the differences between the 2G and the 1G tangential velocity dispersion (see Fig. \ref{fig:vdispratio}): the 2G tangential velocity dispersion is smaller than that of the 1G population. The radial velocity dispersions of the two populations are similar.
  For clusters  that have lost a significant fraction of their mass due to two-body relaxation and are in the advanced stages of their evolution both the 1G and the 2G populations have  isotropic velocity distributions at all clustercentric distances (see Figs. \ref{fig:anisotropy1}-\ref{fig:anisotropyamod2} for the various cases).
\item We have studied the evolution of the 1G and the 2G populations toward energy equipartition and explored the implications of the different initial structural properties of the two populations for their evolution toward energy equipartition.   For dynamically old clusters which have reached spatial and kinematic mixing, the degree of energy equipartition of the 1G is similar to that of the 2G populations (see Fig. \ref{fig:equipmeq} upper panel). For clusters in less advanced stages of their evolution, however, we find that, with the exception of the inner regions where the two populations have similar degrees of energy equipartition, the 2G population is closer to energy equipartition than the 1G population (see Fig. \ref{fig:equipmeq} lower panel).
\end{itemize}

In this paper we have presented an overview of the main dynamical processes driving the evolution of multiple-population clusters, explored the evolutionary paths followed by the 1G and 2G structural and kinematic properties, and studied their relationship with various dynamical parameters. In future works we will expand the range of initial conditions considered and further explore the evolution of the dynamical properties of multiple-population clusters and their dependence on the initial conditions.

\section*{Acknowledgements}
EV acknowledges support from NSF grant AST-2009193. JH was supported by Basic Science Research Program through the National Research Foundation of Korea (NRF) funded by the Ministry of Education (No. 2020R1I1A1A01051827). MG and AH were partially supported by the Polish National Science Center (NCN) through the grant UMO-2016/23/B/ST9/02732. AH is also supported by the Polish National Science Center grant Maestro 2018/30/A/ST9/00050.

\section*{Data availability statement}
The data presented in this article may be shared on reasonable request to the corresponding author.






\bsp	
\label{lastpage}
\end{document}